\documentclass[sigconf]{acmart}
\AtBeginDocument{%
  }

\let\sigproof\proof\let\proof\relax
\let\sigendproof\endproof\let\endproof\relax
\usepackage[utf8]{inputenc} 
\usepackage[T1]{fontenc}
\usepackage{url}
\usepackage{lipsum}
\usepackage{ifthen}
\usepackage{bbm}
\usepackage{listings}
\usepackage{hyperref}
\usepackage{booktabs}
\usepackage{nicematrix}
\usepackage{amsmath}[cmex10] 
\usepackage{enumitem}
\usepackage{amsfonts,bm,graphicx,latexsym,float}
\usepackage{rotating,setspace,latexsym,epsf,color,subcaption}
\usepackage{dblfloatfix}
\usepackage{comment,caption}
\usepackage{soul}
\usepackage{tcolorbox}
\usepackage{xcolor}
\usepackage{tabularx}
\usepackage{pifont}
\usepackage{nicematrix}
\usepackage{multirow} 
\usepackage{array} % For better column formatting
\usepackage{lscape} % Optional: Use for landscape orientation if needed
\usepackage{longtable}
\usepackage{makecell}
\usepackage{comment}
\usepackage{longtable}
\usepackage{array}
\usepackage{ragged2e}
\usepackage{setspace}
\usepackage{longtable}
\usepackage{array}
\usepackage{ragged2e}
\usepackage{tcolorbox}
\usepackage{fvextra}
\usepackage{float}

\definecolor{customteal}{RGB}{105, 179, 162}  % #69B3A2
\definecolor{custompeach}{RGB}{244, 162, 97}  % #F4A261
\newcommand{\cmark}{\ding{51}} % 
\newcommand{\xmark}{\ding{55}} % 

\newcommand{\xhdr}[1]{{\vspace{1pt}\noindent\bfseries #1}.}
\newcommand{\jialin}[1]{\textcolor{purple}{(Jialin: #1)}}

\let\proof\sigproof
\let\endproof\sigendproof

\newtheoremstyle{sig}
  {}
  {}
  {\itshape}
  {}
  {\scshape}
  {.}
  {.5em}
  {#1 #2\thmnote{\quad(#3)}}

\definecolor{niceblue}{HTML}{007ED6}

\definecolor{wine}{HTML}{C70039}

\lstdefinestyle{promptstyle}{                           % Tab size
backgroundcolor=\color{gray!5},       % Light gray background
basicstyle=\ttfamily,    % Basic font style
frame=single,                         % Single frame around the code
rulecolor=\color{gray},               % Frame color
breaklines=true,                      % Break long lines
breakindent=0pt,                 % Red arrow at break
keywordstyle=\color{niceblue}\bfseries,   % Keywords in blue bold
commentstyle=\color{green}\itshape,   % Comments in green italic
stringstyle=\color{orange},           % Strings in orange
showstringspaces=false,               % Don't show spaces in strings
tabsize=4,                            % Tab size
captionpos=t,                         % Caption position
morekeywords={Question, Answer, Ground, Truth, Passage, Abstract, Title, Levels, Input, Output, Topics, Website, Prompt, Model, Provided} % Add more keywords if needed
}
\DeclareCaptionFormat{listing}{#1#2#3}
\newcommand{\setlistingname}[1]{%
  \ifthenelse{\equal{#1}{Prompt}}%
    {\renewcommand{\lstlistingname}{Prompt}}%
    {\renewcommand{\lstlistingname}{Example}}%
}

\lstdefinestyle{examplestyle}{                        % Define style for Example
  backgroundcolor=\color{gray!5},                     % Light gray background
  basicstyle=\ttfamily,                               % Basic font style
  frame=single,                                       % Single frame around the code
  rulecolor=\color{gray},                             % Frame color
  breaklines=true,                                    % Break long lines
  breakindent=0pt,                                    % No break indent
  keywordstyle=\color{wine}\bfseries,                 % Keywords in wine red bold
  commentstyle=\color{green}\itshape,                 % Comments in green italic
  stringstyle=\color{orange},                         % Strings in orange
  showstringspaces=false,                             % Don't show spaces in strings
  tabsize=4,                                          % Tab size
  captionpos=t,                                       % Caption position
  morekeywords={Example, Question, Answer, Ground, Truth, Passage, Abstract, Title, Levels, Input, Output, Topics, Website, Model, Provided} % Add more keywords if needed
}

\copyrightyear{2026}
\acmYear{2026}
\setcopyright{cc}
\setcctype{by}
\acmConference[KDD 2026] {Proceedings of the 32nd ACM SIGKDD Conference on Knowledge Discovery and Data Mining V.1}{August 9--13, 2026}{Jeju Island, Republic of Korea.}
\acmBooktitle{Proceedings of the 32nd ACM SIGKDD Conference on Knowledge Discovery and Data Mining V.1 (KDD 2026), August 9--13, 2026, Jeju Island, Republic of Korea}
\acmISBN{979-8-4007-2258-5/2026/08}
\acmDOI{10.1145/3770854.3785679}

\begin{document}
\title{LitBench: A Graph-Centric Large Language Model Benchmarking Tool For Literature Tasks}

\author{Andreas Varvarigos}
\email{andreas.varvarigos@yale.edu}
\affiliation{%
  \institution{Yale University}
  \streetaddress{}
  \city{New Haven}
  \state{CT}
  \country{USA}
  \postcode{06511}
}

\author{Ali Maatouk}
\email{ali.maatouk@yale.edu}
\affiliation{%
  \institution{Yale University}
  \streetaddress{}
  \city{New Haven}
  \state{CT}
  \country{USA}
  \postcode{06511}
}

\author{Jiasheng Zhang}
\email{jiasheng.zhang.jz875@yale.edu}
\affiliation{%
  \institution{Yale University}
  \streetaddress{}
  \city{New Haven}
  \state{CT}
  \country{USA}
  \postcode{06511}
}

\author{Ngoc Bui}
\email{ngoc.bui@yale.edu}
\affiliation{%
  \institution{Yale University}
  \streetaddress{}
  \city{New Haven}
  \state{CT}
  \country{USA}
  \postcode{06511}
}

\author{Jialin Chen}
\email{jialin.chen@yale.edu}
\affiliation{%
  \institution{Yale University}
  \streetaddress{}
  \city{New Haven}
  \state{CT}
  \country{USA}
  \postcode{06511}
}

\author{Leandros Tassiulas}
\email{leandros.tassiulas@yale.edu}
\affiliation{%
  \institution{Yale University}
  \streetaddress{}
  \city{New Haven}
  \state{CT}
  \country{USA}
  \postcode{06511}
}

\author{Rex Ying}
\email{rex.ying@yale.edu}
\affiliation{%
  \institution{Yale University}
  \streetaddress{}
  \city{New Haven}
  \state{CT}
  \country{USA}
  \postcode{06511}
}

\renewcommand{\shortauthors}{Andreas Varvarigos et al.}

\begin{abstract}
While large language models (LLMs) have become the de facto framework for literature-related tasks, they still struggle to function as domain-specific literature agents due to their inability to connect pieces of knowledge and reason across domain-specific contexts, terminologies, and nomenclatures. This challenge underscores the need for a tool that facilitates such domain-specific adaptation and enables rigorous benchmarking across literature tasks. To that end, we introduce LitBench, a benchmarking tool designed to enable the development and evaluation of domain-specific LLMs tailored to literature-related tasks. At its core, LitBench uses a data curation process that generates domain-specific literature sub-graphs and constructs training and evaluation datasets based on the textual attributes of the resulting nodes and edges. The tool is designed for flexibility, supporting the curation of literature graphs across any domain chosen by the user, whether high-level fields or specialized interdisciplinary areas. In addition to dataset curation, LitBench defines a comprehensive suite of literature tasks, ranging from node and edge level analyses to advanced applications such as related work generation. These tasks enable LLMs to internalize domain-specific knowledge and relationships embedded in the curated graph during training, while also supporting rigorous evaluation of model performance. Our results show that small domain-specific LLMs trained and evaluated on LitBench datasets achieve competitive performance compared to state-of-the-art models like GPT-4o and DeepSeek-R1. To enhance accessibility and ease of use, we open-source the tool along with an AI agent tool that streamlines data curation, model training, and evaluation.

%thus enabling LLMs to internalize domain-specific knowledge and relationships embedded in the curated graph during training.

\end{abstract}



\maketitle

\section{Introduction}
\label{sec:intro}
Despite the widespread adoption of general-purposes LLMs such as GPT-4o~\citep{achiam2023gpt}, Gemini~\citep{team2024gemini}, DeepSeek~\citep{liu2024deepseek}, domain-specific LLMs are gaining traction for their ability to tackle targeted tasks that require deep, specialized domain expertise, such as biomedical research, law, finance, or various interdisciplinary fields~\citep{ling2023domain}. This growing preference stems from a mismatch between the vast but often general-purpose training corpora used by major LLMs and the specialized terminologies, structured domain knowledge, and complex information networks in the form of \emph{literature graphs} that characterize these specialized domains.

Given this context, recent research efforts have led to the development of specialized LLMs across various domains, including BioMistral~\citep{labrak2024biomistral}, PMC-LLaMA~\citep{wu2024pmc}, FinGPT~\citep{liu2023fingpt}, and  ClinicalGPT~\citep{wang2023clinicalgpt}. 
\begin{comment}
ChatLAW~\citep{cui2023chatlaw}
\end{comment}
These efforts highlight that simply fine-tuning on domain-specific text can significantly improve model performance on tasks like named entity recognition, classification, or short-form question answering. Yet, most of these approaches rely on static corpora and do not extensively utilize or represent the rich structural relationships between pieces of knowledge manifested in domain-specific literature graphs. As a result, a gap remains between the general abilities of LLMs and the depth of understanding required for high-level literature-based tasks such as synthesizing related work sections or exploring novel research questions in niche interdisciplinary domains~\citep{zhang2024litfm}.

To that end, recent research has turned toward integrating knowledge graphs and citation networks into LLM training and fine-tuning pipelines~\citep{tang2024graphgpt, zhang2024litfm}. Although large-scale citation graphs, such as the Microsoft Academic Graph (MAG)~\cite{sinha2015overview}, Semantic Scholar’s Open Research Corpus (S2ORC)~\citep{lo2019s2orc}, Arxiv~\citep{saier2023unarxive}, and OpenAlex~\citep{priem2022openalex}, provide extensive bibliographic and citation relationships, they are often incomplete. These resources frequently miss crucial textual components, including introductions, related work sections, or the specific citation sentences that articulate the relationships among papers. Moreover, extracting and curating relevant subsets from these vast datasets for specific sub-domains typically requires significant manual effort in selecting, cleaning, and transforming the literature into a format suitable for training and evaluation. Such challenges are further amplified in specialized tasks like related work generation, which demand data that is both comprehensive and nuanced.

In parallel, a variety of benchmarks have emerged to evaluate domain-specific LLM capabilities in scientific contexts~\citep{jain2020scirex, tsatsaronis2015overview} 
\begin{comment}
    \citep{jain2020scirex, tsatsaronis2015overview, nentidis2023overview}
\end{comment}
as well as other specialized areas like cybersecurity, clinical medicine, and biomedical research. Despite their contributions, these benchmarks frequently focus on sentence-level or short-passage tasks, lack fully automated data curation workflows, or do not test higher-level narrative generation skills such as compiling a coherent literature review. This gap limits their effectiveness in testing whether LLMs can truly act as domain experts in literature, underscoring the need for a systematic framework that not only streamlines data curation and domain adaptation but also evaluates model performance across a comprehensive suite of literature-related tasks, including understanding and synthesizing interrelated material.

With this in mind, in this paper, we introduce \textit{LitBench}, a benchmarking tool designed to create, train, and evaluate LLMs capable of effectively performing literature tasks in user-defined specialized domains. At its core, LitBench employs an automated data curation pipeline that constructs \textit{domain-specific literature subgraphs} from research papers, sourced from arXiv due to its open-access nature and permissive mining policies\footnote{Our tool can also process offline user-defined research paper datasets.}. To achieve this, the first step involves augmenting arXiv's default taxonomy by generating nine topics at varying levels of abstraction for each paper. This multi-level hierarchical representation enhances topic granularity and enables the construction of domain-specific subgraphs, regardless of how broad or specialized the user-specified domain may be.

In the second stage of LitBench, a topic-based retriever powered by an encoder LLM leverages the hierarchical topic representations to identify material relevant to the user's target domain. Combined with a custom-built LaTeX parser, we construct a rich domain-specific citation graph that captures essential textual elements, such as citation sentences and related work sections, critical for training and evaluating LLMs on literature-related tasks. This graph serves as the foundation for generating both instruction-tuning and benchmarking datasets that accurately capture the vocabulary, terminologies, and contextual relationships unique to the specified domain, ultimately enabling LLMs to perform complex literature synthesis tasks with greater domain expertise.

The graph-centric approach of LitBench not only supports the flexible curation of both broad and niche subdomains, thus eliminating the need to create individual benchmarks for various domains, but also embeds relational context, such as citation links and topical similarities, directly into model training through our instruction fine-tuning datasets. Moreover, this focus on a specialized graph compels the model to learn to distinguish between similar yet unrelated pieces of knowledge, which is less likely to happen when training on broad, unspecialized graphs. As a result, LLMs fine-tuned with LitBench internalize domain-specific language and literature structures and achieve performance that rivals state-of-the-art models. Furthermore, LitBench integrates a comprehensive suite of literature-related tasks, ranging from simple tasks like title and abstract generation to advanced tasks such as related work generation. This multi-faceted evaluation framework provides a rigorous assessment of model performance across diverse literature tasks.

\noindent\textbf{Contributions.} In the following, we summarize the contributions of our paper. 
\begin{itemize}[leftmargin=*]
    \item We introduce a curation tool that constructs domain-specific subgraphs from research papers by constructing a multi-level hierarchical representation of paper concepts with a custom-built LaTeX parser. Unlike existing datasets~\citep{sinha2015overview}, 
    \begin{comment}
        \citep{sinha2015overview, lo2019s2orc, saier2023unarxive, priem2022openalex}
    \end{comment}
    and as illustrated in Table~\ref{comparison_table} and Appendix \ref{appendix:datasetssota}, our approach incorporates rich textual elements—such as citation sentences, related work, and introductions—to enhance LLMs' understanding of literature-related tasks.
    \item We propose an efficient retriever that leverages the hierarchical topic structure to generate flexible training and benchmarking datasets for domain-specific LLMs, supporting literature-related tasks such as abstract generation, title suggestion, and related work synthesis.
    \item We open-source our tools\footnote{\url{https://github.com/varvarigos/LitBench}}, complete with a user-friendly GUI, enabling users to easily create customized datasets and train specialized LLMs for any domain of interest, regardless of how broad or niche.
\end{itemize}

\begin{figure*}[h]
    \centering
    \includegraphics[width = \linewidth]{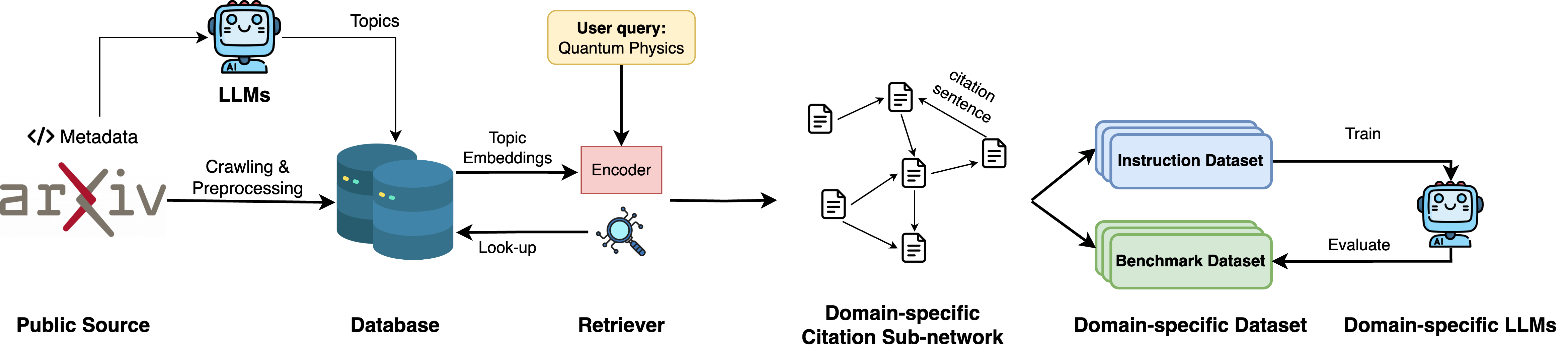}
    \vspace{3pt}
    \caption{Overall pipeline of LitBench. The framework begins with arXiv-sourced metadata, from which we extract textual content using a specialized LaTeX parser. A topic-based retriever, guided by user queries (e.g., “Quantum Physics”), then constructs domain-specific citation graphs. These graphs are subsequently transformed into instruction and benchmarking datasets, enabling LLMs to be trained and evaluated on specialized, literature-related tasks.}
    \label{fig:overall}
    \vspace{-10pt}
\end{figure*}

\section{Related Work}
\begin{table}[]
\centering
\caption{Differences in graph attributes in current datasets.}
\label{comparison_table}
\resizebox{0.99\linewidth}{!}{
\begin{tabular}{@{}lcccc@{}}
\toprule
\textbf{Datasets}     & \textbf{Titles \& Abstracts}          & \textbf{Citation Sentences} & \textbf{Intro \& Related Work} & \textbf{Concepts}  \\ \midrule
MAG        &  \cmark     & \xmark     &  \xmark                       &     \xmark                                \\

ArXiv        &  \cmark     & \xmark     &  \xmark                       &     \xmark                                \\
OpenAlex     &  \cmark       &  \xmark            &   \xmark                          &      \cmark                                        \\

S2ORC        &  \cmark     & \cmark     &  \cmark                       &     \xmark                                \\

\textbf{LitBench} &  \cmark        &  \cmark               &   \cmark                   &           \cmark                           \\ \bottomrule
\end{tabular}}
\vspace{-10pt}
\end{table}
\textbf{Domain Specification and Adaptation of LLMs.} Building LLMs that are effective within specific domains—ranging from scientific literature to legal text—has been a growing area of research. Early work in this direction focused on pre-training BERT-based architectures on specialized corpora, such as biomedical (BioBERT) and scientific domains (SciBERT) to capture domain-relevant terminologies and linguistic patterns \cite{beltagy2019scibert}. 
\begin{comment}
    \cite{beltagy2019scibert, lee2020biobert}
\end{comment}
Subsequent efforts extended to even more specialized models like ClinicalBERT \cite{huang2019clinicalbert}, ClimateBERT~\cite{webersinke2021climatebert} and LegalBERT \cite{chalkidis2020legal}, further underscoring the importance of tailoring models to domain-specific text distributions. This trend continues to expand in the era of large models in domains such as medicine \cite{labrak2024biomistralcollectionopensourcepretrained},
\begin{comment}\cite{labrak2024biomistralcollectionopensourcepretrained, singhal2025toward, singhal2023large, labrak2024biomistral, wu2024pmc, wang2023clinicalgpt}\end{comment}
law \cite{colombo2024saullm7bpioneeringlargelanguage},
\begin{comment}
\cite{colombo2024saullm7bpioneeringlargelanguage, cui2023chatlaw}
\end{comment}
finance \cite{xie2023efficientcontinualpretrainingbuilding}.
\begin{comment}
\cite{xie2023efficientcontinualpretrainingbuilding, wu2023bloomberggpt, liu2023fingpt}
\end{comment}
However, these specialized models typically rely on static corpora and often do not incorporate explicit knowledge or citation graphs, limiting their ability to understand complex relationships among concepts within a domain’s literature \cite{li2024synergizing}.

\textbf{Literature Graph Curation and Knowledge Extraction.} The idea of using structured knowledge to enrich language models has led to the development of various knowledge-enhanced frameworks. Prior studies have leveraged external knowledge bases (KBs) such as Wikidata or domain-specific ontologies to improve model understanding and generation tasks \cite{pan2024unifying}. 
\begin{comment}
    \cite{pan2024unifying, sun2023think}
\end{comment}
In the scientific domain, resources like the Microsoft Academic Graph (MAG) \cite{sinha2015overview}, Semantic Scholar’s Open Research Corpus (S2ORC) \cite{lo2019s2orc}, Arxiv~\citep{saier2023unarxive}, and OpenAlex \cite{priem2022openalex} have been used to create large citation networks and bibliographic datasets for downstream tasks such as citation recommendation, article classification, and summarization \cite{cohan2020specter}. While these datasets capture paper-level and citation-level relationships, they often require extensive manual filtering or adaptation to support the training of domain-focused LLMs. Moreover, existing approaches rarely provide a straightforward mechanism to curate and transform these large-scale literature graphs into specialized sub-graphs that align with particular interdisciplinary or niche domains.

\noindent\textbf{Benchmarks for Scientific and Domain-Specific Tasks.} There is a growing body of benchmarks designed to evaluate LLM performance on scientific text. Notable examples are SciREX \cite{jain2020scirex}, ACL-ARC \cite{jurgens2018measuring} and SCICITE \cite{cohan2019structural} in general scientific domain and BioASQ \cite{tsatsaronis2015overview} 
\begin{comment}
    \cite{tsatsaronis2015overview, nentidis2023overview}
\end{comment}
in biomedical domain, CyberMetric~\citep{tihanyi2024cybermetric} in cybersecurity domain, MEDIC in~\citep{kanithi2024medic, luo2024large} and MedBench~\citep{cai2024medbench} in  medical domain. While these domain-specific benchmarks offer valuable insights, they generally require extensive manual data curation. In contrast, we focus on literature-related tasks and aim to automate the data curation process, thereby streamlining the creation of specialized benchmark datasets for domain-specific LLM evaluation.
% . However, most of these remain limited to sentence-level or short-passage tasks, lacking comprehensive tasks such as generating complete related work sections or adapting to niche sub-domains. Recently, several benchmark datasets have been introduced to evaluate LLMs in more specialized areas like scientific proficiencyon.

\section{DATASET COLLECTION \& Curation}

\subsection{Overview}
%Our dataset collection and curation process automatically generates rich, domain-specific literature graphs that underpin our benchmark framework (see Figure~\ref{fig:overall}). We begin by crawling arXiv papers metadata from Kaggle\footnote{\url{https://www.kaggle.com/datasets/Cornell-University/arxiv}}, resulting in a collection of approximately 2.3M papers information. The metadata of each paper is then used
%\ref{appendix:prompts}

%With the meta-

%Since all documents are based on LaTeX sources, we parse key sections—such as the introduction and related work—to extract relevant information. Following the approach in~\citep{saier2023unarxive}, we adopt a JSON structure (see Table~\ref{tab:}) for each paper's metadata, including the title, abstract, introduction, license, discipline, and more. Additionally, for each paper, we extract nine topics from the abstract and title by prompting a powerful LLM (e.g., GPT-4o). These topics are organized into three levels of abstraction to facilitate efficient retrieval of domain-specific papers.
%For retrieval, given an input query specifying the target domain for LLM training, we propose a topic-based retriever that identifies relevant papers based on the embedding similarity between the query and extracted concepts. The retrieved documents form a domain-specific citation sub-network, which is then used to construct both instruction and benchmark datasets for training and evaluating our domain-specific LLMs. 

Our dataset collection and curation process automatically generates rich, domain-specific literature graphs that form the foundation of our benchmark framework (see Figure~\ref{fig:overall}). We begin by collecting arXiv paper metadata from the Kaggle dataset that is updated monthly\footnote{\url{https://www.kaggle.com/datasets/Cornell-University/arxiv}}, resulting in a corpus of approximately 2.3 million papers. Using this metadata, we prompt a powerful LLM to extract nine topics/concepts from the abstract and title of each paper. These concepts are organized into three levels of abstraction, thus enabling versatile retrieval of relevant papers for any area or niche domain specified by the user. For retrieval, given an input query specifying the desired domain, we employ a topic-based retriever that identifies relevant papers based on the embedding similarity between the query and the extracted concepts. The identified papers are then crawled in LaTeX format, parsed to extract citations, and processed to isolate key sections, such as the introduction and related work. This process constructs a domain-specific citation sub-network, which is used to create both instruction and benchmark datasets for training and evaluating domain-specific LLMs.

\subsection{Concepts Curation}

%Our process begins with crawling the arXiv papers metadata snapshot from Kaggle, which includes approximately 2.3 million papers. This metadata contains a variety of information for each paper, such as the title and abstract. To enable the creation of domain-specific subgraphs, our first objective is to align the information in these metadata with user queries. To achieve this, we utilize a powerful LLM (in our case, Meta-LLaMA-3.1-70B-Instruct) and provide it with the title and abstract of each paper. The model is prompted three times, each with a distinct prompt designed to extract topics at varying levels of abstraction. Level 1 represents the highest level of abstraction (e.g., broad fields like Computer Science), while Level 3 captures more fine-grained, specific topics (e.g., methodologies or techniques). The prompts used for this process are provided in Appendix \ref{appendix:prompts}.

%\textbf{Motivation.} 
%The motivation behind using multiple abstraction levels is to create representative topics/concepts in natural language that cater to diverse user needs. For instance, a paper on LLMs might have a high-level topic like "Computer Science" and a low-level topic like "Transformer-based architectures." This hierarchical representation allows users to retrieve papers at the desired level of specificity, whether they are interested in broad domains or niche methodologies. Examples of these extracted concepts are provided in Appendix \ref{appendix:concepts}. To facilitate access and reuse, we have made these concepts available on Huggingface\footnote{\url{https://huggingface.co/datasets/AliMaatouk/arXiv_Topics}}.

\xhdr{Motivation} The motivation behind creating concepts for the papers with multiple abstraction levels is to create representative topics/concepts in natural language that cater to diverse user needs. For instance, a paper on LLMs might have a high-level topic like "Computer Science" and a low-level topic like "Transformer-based architectures." This hierarchical representation allows users to retrieve papers at the desired level of specificity, whether they are interested in broad domains or niche methodologies.

To create such concepts and enable our LitBench framework to create domain-specific subgraphs, our process begins with crawling the arXiv papers metadata snapshot from Kaggle, which includes approximately 2.3 million papers. This metadata contains a variety of information for each paper, such as the title and abstract. To align the information in these metadata with user queries, we utilize a powerful LLM (in our case, Meta-LLaMA-3.1-70B-Instruct) and provide it with the title and abstract of each paper. The model is prompted three times, each with a distinct prompt designed to extract topics at varying levels of abstraction. Level 1 represents the highest level of abstraction (e.g., broad fields like Computer Science), while Level 3 captures more fine-grained, specific topics (e.g., methodologies or techniques). The prompts used for this process, along with examples of these extracted concepts, are provided in Appendix \ref{appendix:prompts} and \ref{appendix:concepts} respectively. To facilitate access and reuse, we have made these concepts available on Huggingface\footnote{\url{https://huggingface.co/datasets/AliMaatouk/arXiv_Topics}}.

%combine the arXiv metadata  and Semantic Scholar citations information \cite{kinney2023semanticscholaropendata} to create a citation graph $\mathcal{G}=(V,E)$. Here, $V$ represents the arXiv papers and $E$ represents citation relationships between papers. In this graph, each node's attributes include its title and abstract. Next, we utilize Meta-LLama-3.1-70B-Instruct and provide it with the title and abstract of each node $\mathcal{G}$. The model is prompted three times, each with a different prompt aimed at extracting topics at varying levels of abstraction. This process generates three graphs $\mathcal{G}^{(i)}=(V^{(i)},E)$ for $i=1,2,3$, where $V^{(i)}$ is the set of papers with attributes being its level $i$ topics. In this context, level 1 represents the highest level of abstraction, while level 3 represents more fine-grained, specific topics. The edge information remains consistent with $\mathcal{G}$. The prompts utilized within our framework are provided in Appendix \ref{appendix:prompts}.

\subsection{Concept-based Retriever}\label{sec:retriever}
The objective of our retriever is to construct a citation subgraph most relevant to the user's topic of interest. Given an input query $q$ and a set of papers $V$, where each paper has attributes such as title $p_t$, abstract $p_a$, and our newly devised concepts attributes $p^i_c$ at varying abstraction levels $i=1,2,3$, we aim to retrieve a subset of papers $V^* \subseteq V$ that are most pertinent to $q$.

The conventional approach retrieves relevant papers by embedding the query into a vector space and calculating the relative distance between the query embedding and the embeddings of papers' titles or abstracts. Let $\phi(\cdot)$ be a pretrained encoder such as BERT \citep{devlin2018bert}. The set of retrieved papers $V^*$ is typically defined as:
\begin{equation}
V^* = \mathrm{TopK}_{p \in V} ~\mathrm{sim}(\phi(q), \phi(p_a)),
\end{equation}
where $\mathrm{sim}(\cdot, \cdot)$ is a similarity function. However, this method achieves suboptimal retrieval results because abstracts and titles often contain redundant information and are not well-aligned with user queries, especially when queries span different abstraction levels. We validate this hypothesis experimentally in Section~\ref{sec:experiment}.

To address this, we leverage the multi-level concept attributes we generated to refine the retrieval process. Instead of relying solely on abstracts or titles, we embed the concepts and compute similarity scores between the query embedding and the average embedding of the concept attributes:
\begin{equation}
V^\star = \mathrm{TopK}_{p \in V} ~\mathrm{sim}\left(\phi(q), \frac{1}{3}\sum_{i=1}^{3} \phi(p^i_c)\right),
\end{equation}
where $\mathrm{sim}(\cdot)$ is chosen as the cosine similarity. This refined approach enables our retriever to better capture the nuances of the user's query, resulting in a more relevant citation subgraph tailored to the intended domain of interest. Similarly, to facilitate reuse, we have made the embeddings of these concepts using BGE-large, a retrieval-optimized encoder model\footnote{\url{https://huggingface.co/BAAI/bge-large-en-v1.5}}, available on HuggingFace\footnote{\url{https://huggingface.co/datasets/AliMaatouk/arXiv-Topics-Embeddings}}. The choice of this embedding model is motivated by its strong performance for its size, as evidenced by its popularity on Hugging Face, and the fact that it has been trained using contrastive learning on datasets that include scientific papers.

\subsection{Sub-Graph Construction}
%With the relevant papers set $V^*$ identified, we pick top-k papers that are most relevant to the query of the user. Typically, k is chosen according to the domain of interest with k being smaller as the domain being more niche to ensure the retrieval accuracy remains high. We analyze later in Section \ref{sec:experiment} the effect of $k$ and how big the graph needs to be to inject well the information of the graph into the LLM. 

%With these papers identified, we proceed with crawling the LaTex sources of these papaers. The crawling of Latex sources requires a principled processing approach in order to not only process and clean the data, that involves flattening, de-macroing 

%More importantly, our processing platform identify citation links among papers, and their citing sentence, along with key attributes such as related works and introduction, thus enriching the datasets and providing additional information compared to datasets as ogbn-arxiv-TA \cite{yan2023comprehensive} and  S2ORC \cite{lo2019s2orc}, which will allow us to internalize more information of the graph in the LLM. The details of this processing and curation process are provided in Appendix \ref{sec:appendix_cleaning}. By doing so, we end up with a domain $G = (V^*, E^*)$ representing the domain-specific subgraph on the basis, containing a list of attributes: concepts, title, abstract, introduction, and related work along with citation sentence along each citation.  

After calculating the similarity between the query and each paper's concept attributes, we select the top-k papers most pertinent to the user's query. The value of k is chosen based on the domain of interest, with smaller k values used for more niche domains to ensure high retrieval accuracy. We analyze the impact of k and the required size of the subgraph for effective knowledge injection into the LLM in Section \ref{sec:experiment}. Once the relevant papers are identified, we proceed to crawl their LaTeX sources. To this end, we develop a principled processing approach to clean and structure the data, which includes flattening the LaTeX files, de-macroing custom commands, and extracting key sections.

More importantly, our processing pipeline identifies citation links between papers, extracts citing sentences, and enriches the dataset with key attributes such as related work and introduction sections. This provides significantly more information compared to existing datasets like ogbn-arxiv-TA \cite{yan2023comprehensive} and S2ORC \cite{lo2019s2orc}, enabling the LLM to internalize a richer representation of the graph. The details of this processing and curation process are provided in Appendix \ref{sec:appendix_cleaning}. By the end of this process, we obtain a domain-specific subgraph $G = (V^*, E^*)$ where each node (paper) contains attributes such as concepts, title, abstract, introduction, related work, and citation sentences associated with each edge (citation link). This enriched representation forms the foundation for training and evaluating domain-specific LLMs.
\subsection{Multi-Instruction Graph Internalization}

With the subgraph $G = (V^*, E^*)$ constructed, we discuss in this section the process of devising new instruction and benchmark datasets using this retrieved domain-specific subgraph. To facilitate training domain-specific LLMs, we create an instruction fine-tuning dataset that leverages both node-level and edge-level information to capture specialized knowledge and inter-paper relationships.

For node-level tasks, we design four tasks: \textit{Title Generation}, \textit{Abstract Completion}, \textit{Related Work Generation}, and \textit{Introduction to Abstract}. These tasks use paper titles, abstracts, and introductions to infuse domain-specific context. By learning to perform these tasks, the model deepens its understanding of the scientific content within this domain. For edge-level tasks, we introduce \textit{Paper Recommendation}, \textit{Citation Link Prediction}, and \textit{Citation Sentence Generation}. These tasks train the model to discern the relevance and connections among papers, particularly in the presence of hard negatives due to the highly specialized nature of the graph, while also learning to generate appropriate citation sentences. This enables it to understand how citations are integrated into scholarly writing—ultimately preparing the model to handle complex literature-related queries like related work generation. Benchmark datasets are constructed similarly to the instruction datasets, but on the hold-out citation sub-graph. For more details on the creation of these datasets, we refer the readers to Appendix \ref{appendix:datasetprep}.

Beyond this, we also define in our benchmarking dataset advanced literature tasks beyond the graph's attributes, such as related work generation and identification of the most influential papers. Compared to the training tasks, these aim to push the literature tasks beyond the attributes and test how well models perform in more advanced tasks.

\begin{figure*}[t]
\centering
\begin{minipage}{0.33\textwidth}
  \centering
  \begin{figure}[H]
    \centering
\includegraphics[width=.99\linewidth]{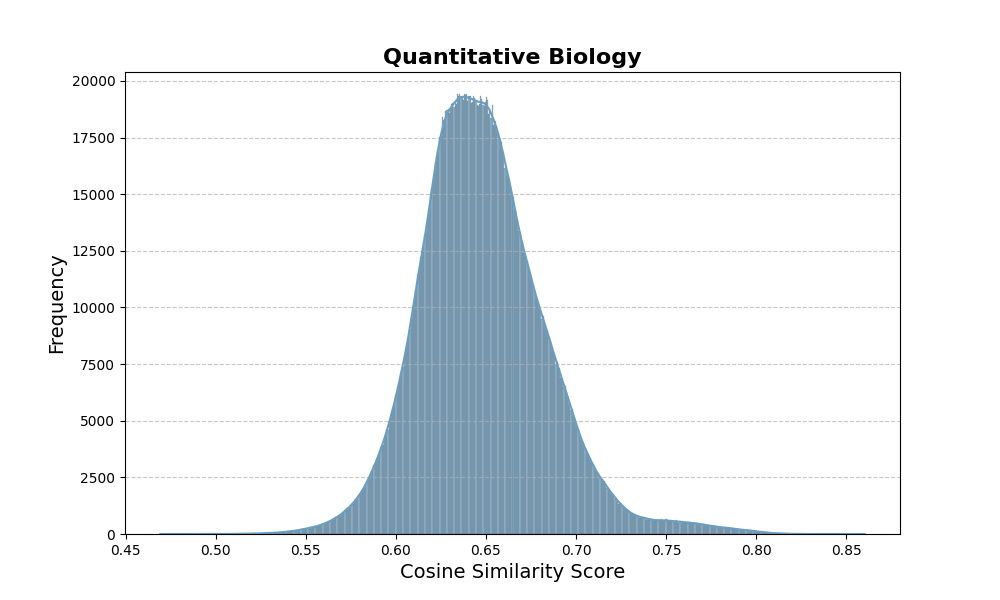}
  \end{figure}
\end{minipage}%
\hfill
\begin{minipage}{0.33\textwidth}
  \centering
  \begin{figure}[H]
    \centering
\includegraphics[width=.96\linewidth]{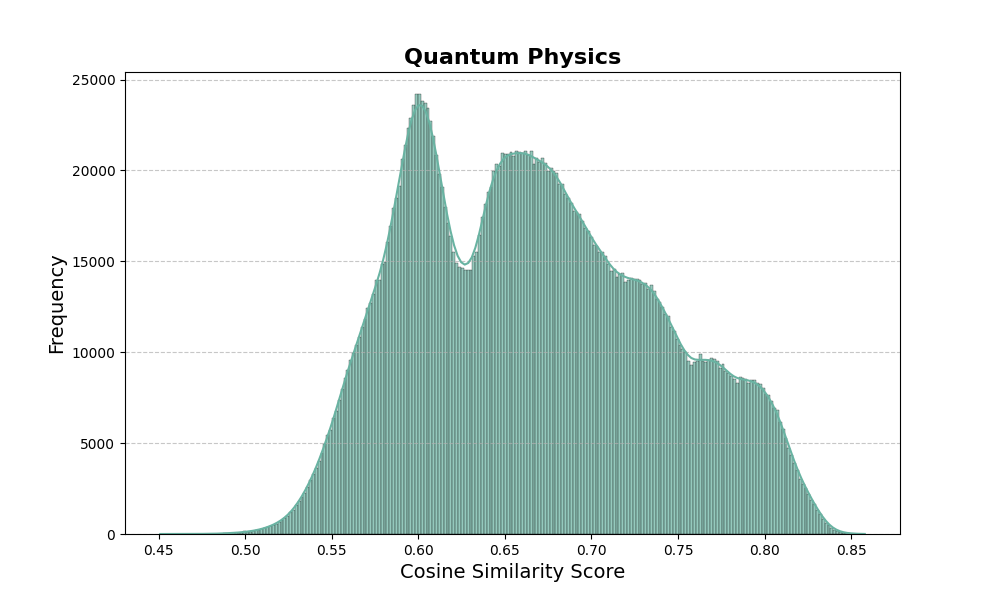}
  \end{figure}
\end{minipage}
\begin{minipage}{0.33\textwidth}
  \centering
  \begin{figure}[H]
    \centering
\includegraphics[width=.99\linewidth]{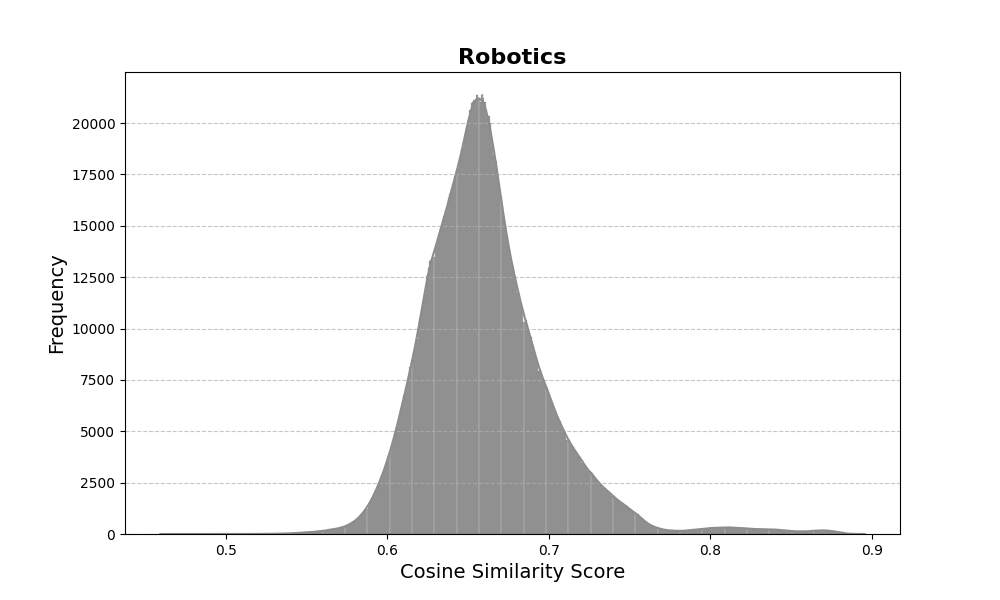}
  \end{figure}
\end{minipage}%
\hfill
\caption{Cosine similarity histograms across the three considered domains: quantitative biology, quantum physics, and robotics.}
\label{fig:tokens}
\vspace{-10pt}
\end{figure*}

\subsection{LitBench Graphical User Interface}
\begin{figure}[t]
    \centering
    \includegraphics[width=0.99\linewidth]{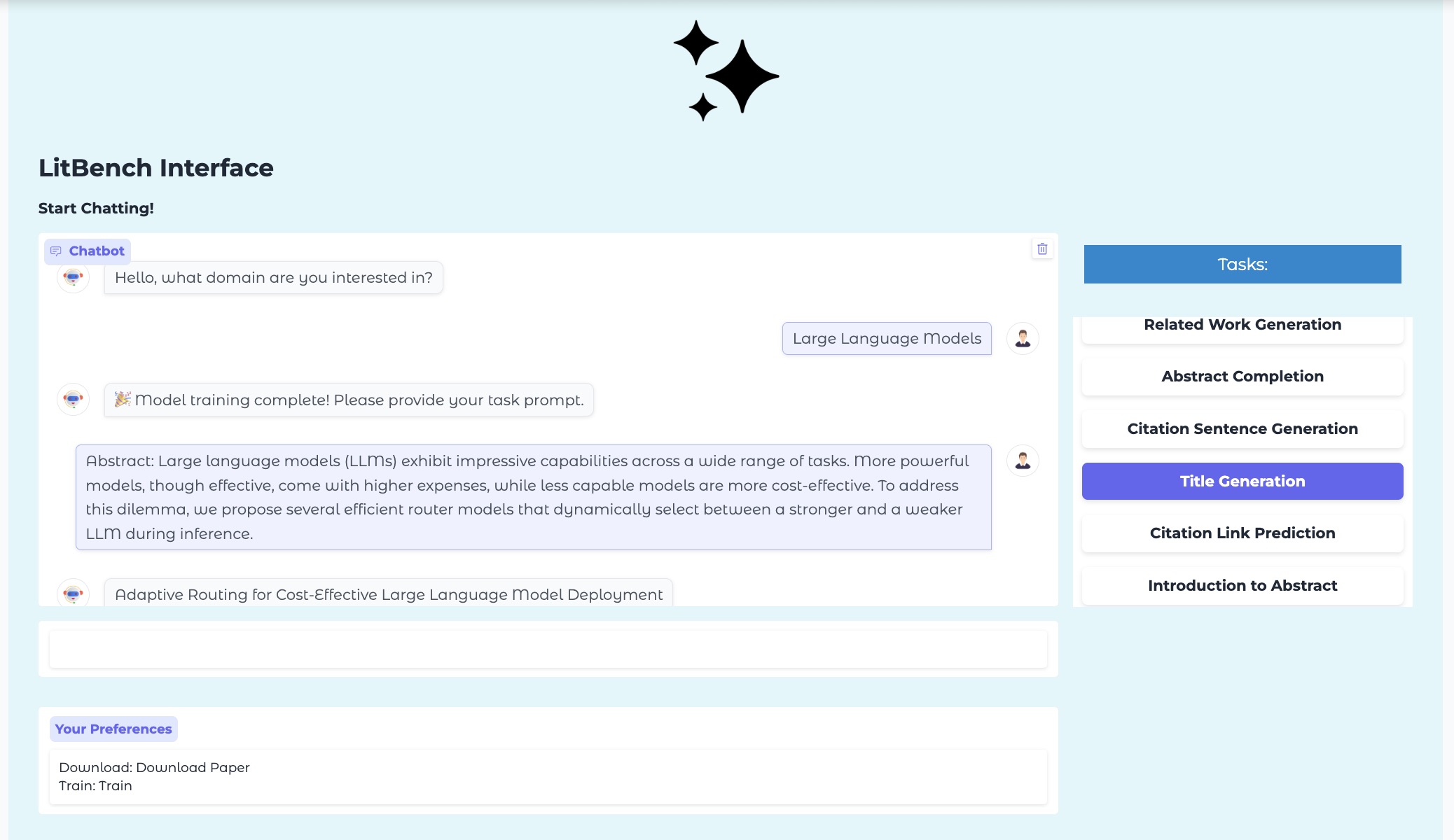}
    \caption{The interactive GUI of the LitBench tool.}
    \label{fig:litbench_interface}
    \vspace{-15pt}
\end{figure}

To simplify the use of LitBench, we developed a GUI tool that leverages the LitBench framework to assist with literature-related tasks across any user-specified topic, whether broad or highly specialized. The tool dynamically generates LitBench datasets for any desired research domain by automating the retrieval and processing of relevant literature. It also facilitates fine-tuning LLMs on LitBench, enabling models to adapt to specific research areas by training on the instruction datasets derived from LitBench. A visual representation of the tool is provided in Fig. \ref{fig:litbench_interface}.

Built using Gradio, the tool offers users two implementation paths: constructing the graph from scratch by downloading and processing arXiv papers locally, or leveraging pre-processed datasets to reduce computational overhead. Additionally, users can choose to either fine-tune an LLM on LitBench tasks or benchmark an already trained model. During benchmarking, users can perform bulk evaluations across all LitBench literature tasks or focus on a specific task of interest. Upon task selection, the system maps the chosen task to its corresponding prompt processing and inference script, ensuring the model generates relevant and accurate responses tailored to the user's needs.

\section{Benchmarking and Evaluation}\label{sec:experiment}

In this section, the evaluation focuses on three main objectives. First, we show how LitBench enables the creation of high-performing domain-specific models that outperform SOTA models. This demonstration is achieved by benchmarking models of various sizes and architectures against those fine-tuned on instruction datasets derived from LitBench across literature tasks. Second, we highlight LitBench's ability to enable the creation of highly specialized domain-specific experts with extreme flexibility while extracting superior performance compared to non-specialized models. Finally, we conduct ablation studies to establish best practices for maximizing model performance using LitBench. These studies explore key factors such as the required size of the subgraph, the impact of warm-starting with pretraining data, and other critical aspects.

\subsection{Experimental Settings \& Datasets}
\xhdr{Models considered}
Throughout our evaluations, we consider a diverse range of open-source models, spanning from 1B to 8B parameters, as baselines. Additionally, we include SOTA closed-source models, specifically GPT-4o and DeepSeek-R1, to provide a comprehensive comparison of performance across both open and proprietary models. We also evaluate a family of LLMs, ranging from 1B to 8B parameters, fine-tuned on instruction datasets constructed from LitBench across various domains. The fine-tuning details for these models are provided in Appendix \ref{appendix:trainingdetails}-1.

\xhdr{LitBench domains} To showcase LitBench's versatility and flexibility, we focus on multiple domains, most importantly, quantitative biology, robotics, and quantum physics. To curate these datasets, we use the domain names as queries and leverage BGE-large to compute cosine similarity  with the embedded concepts field from our devised arXiv metadata graph.  Fig. \ref{fig:tokens} displays the resulting cosine similarity histograms, which guide the selection of relevant papers. Based on these distributions and computational considerations, we construct domain-specific subgraphs, with detailed statistics provided in Table \ref{tab:datasets}. Using these subgraphs, we then generate literature tasks for both fine-tuning and evaluation, following the methodology described earlier. The training token counts for each domain are included in Table \ref{tab:datasets}. %Complete dataset specifications found in this sections are available in Appendix \ref{appendix:datasets}.

\begin{table}[h!]
\caption{LitBench datasets statistics across multiple domains.}
\centering
\resizebox{0.85\linewidth}{!}{
\begin{NiceTabular}{|c|ccc|} 
\toprule 
\textbf{Datasets}
&\multicolumn{1}{c}{\#Nodes}
&\multicolumn{1}{c}{\#Edges}
&\multicolumn{1}{c}{\#Training Tokens}\\
\midrule 
Quantative Biology  & 11.81k &	13.82k	& 282M  \\
Robotics  & 13.93k	& 20.01k	& 373M 	\\
Quantum Physics & 13.08k &	19.74k &	362M	 \\
\bottomrule
\end{NiceTabular}}
\vspace{-10pt}
\label{tab:datasets}\vspace{-0.3cm}
\end{table}

\begin{table*}[]
\caption{Evaluation of various models on generative tasks across quantitative biology, robotics, and quantum physics.}
\centering
\resizebox{0.8\linewidth}{!}{
\begin{NiceTabular}{c|c|cc|cc|cc|cc} 
\toprule 
\textbf{Models}
&\textbf{Datasets}
&\multicolumn{2}{c}{\textbf{Title Generation}}
&\multicolumn{2}{c}{\textbf{Abstract Completion}}
&\multicolumn{2}{c|}{\makecell{\textbf{Citation Sentence} \\ \textbf{Generation}}}
&\multicolumn{2}{c}{\makecell{\textbf{Introduction} \\ \textbf{to Abstract}}}
\\
\midrule 
&\textbf{Metric}&{Precision}&{F-score}&{Precision}&{F-score}&{Precision}&{F-score}&{Precision}&{F-score}\\
\midrule 
\multirow{3}{*}{Llama-3.2-1B}&Biology &0.7915 &0.8037  &0.8185 &0.8075 &0.7915  &0.8037 &0.8056 &0.7993 \\
 &Robotics  &0.7924 &0.8029  &0.8251 &0.8145 &0.8026  &0.7875 &0.8022  &0.7975\\
 &Quantum &0.7873 &0.7990  &0.8144 &0.8076 &0.8059  &0.7830 &0.7967 &0.7927 \\

\midrule

\multirow{3}{*}{Llama-3.2-1B-Lit}&Biology &0.7719 &0.8073  &0.8392 &0.8393 &0.8261  &0.8216 &0.8224 &0.8317 \\
 &Robotics  &0.8066 &0.8362  &0.8446 &0.8487 &0.8349  &0.8329 &0.8305 &0.8361 \\
 &Quantum &0.7868 &0.8189  &0.8412 &0.8455 &0.8291  &0.8212 &0.8300 &0.8343 \\

\midrule

\multirow{3}{*}{Llama 3.2-3B}&Biology &0.7985 &0.8133  &0.8263 &0.8168 &0.7980  &0.7837 &0.7873 &0.7920 \\
 &Robotics  &0.8035 &0.8179  &0.8362 &0.8282 &0.8033  &0.7902 &0.8044 &0.8020\\
 &Quantum &0.7994 &0.8123  &0.8336 &0.8274 &0.8058  &0.7859 &0.8008 &0.7965 \\

\midrule

\multirow{3}{*}{Llama 3.2-3B-Lit}&Biology &0.7985 &0.8284  &0.8439 &0.8474 &0.8303  &0.8254 &0.8359 &0.8403 \\
 &Robotics  &0.7980 &0.8302  &0.8440 &0.8503 &0.8380  &0.8367 &0.8344 &0.8405 \\
 &Quantum &0.7899 &0.8199  &0.8514 &0.8586 &0.8303  &0.8226 &0.8342 &0.8378 \\

 \midrule

\multirow{3}{*}{Llama-3-8B}&Biology &0.7732 &0.7812  &0.8578 &0.8394 &0.8283  &0.8119 &0.8163 &0.8099 \\
 &Robotics  &0.8378 &0.8408  &0.8409 &0.8223 &0.8099  &0.7941 &0.8538 &0.8462 \\
 &Quantum &0.8328 &0.8349  &0.8496 &0.8337 &0.7983  &0.7777 &0.8479 &0.8404 \\

 \midrule

 \multirow{3}{*}{Llama-3-8B-Lit}&Biology &0.7780 &0.8130  &0.8461 &0.8460 &0.8294  &0.8252 &0.8399 &0.8420 \\
 &Robotics  &0.7864 &0.8194  &0.8398 &0.8482 &0.8374  &0.8367 &0.8464 &0.8500  \\
 &Quantum &0.7938 &0.8216  &0.8546 &0.8632 &0.8263  &0.8207 &0.8391 &0.8421 \\

  \midrule

 \multirow{3}{*}{Mistral 7B}&Biology &0.8319 &0.8448  &0.8719 &0.8506 &0.8255  &0.8046 &0.8440 &0.8305 \\
 &Robotics  &0.8395 &0.8537  &0.8761 &0.8576 &0.8276  &0.8091 &0.8561 &0.8419 \\
 &Quantum &0.8166 &0.8356  &0.8640 &0.8510 &0.8263  &0.8008 &0.8465 &0.8347 \\

   \midrule

 \multirow{3}{*}{Vicuna 7B}&Biology &0.8588 &0.8623  &0.8664 &0.8396 &0.8289  &0.8057 &0.8634 &0.8432 \\
 &Robotics  &0.8731 &0.8774  &0.8708 &0.8468 &0.8317  &0.8110 &0.8641 &0.8461 \\
 &Quantum &0.8512 &0.8542  &0.8717 &0.8508 &0.8274  &0.8006 &0.8580 &0.8412\\

   \midrule

 \multirow{3}{*}{GPT-4o}&Biology &0.8498 &0.8621  &0.8597 &0.8413 &0.8224  &0.8075 & 0.8726 & 0.8378 \\
 &Robotics  &0.8678 &0.8784  &0.8644 &0.8483 &0.8386  &0.8156 & 0.8795 &0.8429 \\
 &Quantum &0.8542 &0.8550  &0.8771 &0.8507 &0.8422  & 0.8092 & 0.8680 & 0.8429 \\

   \midrule

 \multirow{3}{*}{DeepSeek-R1}&Biology &0.7772 &0.8161  &0.7791 &0.8011 &0.8077 &0.7963 &0.8050 &0.8163 \\
 &Robotics  &0.7969 &0.8361  &0.8095 &0.8183 &0.8106  &0.8078 &0.8454 &0.8267 \\
 &Quantum &0.7959 &0.8310  &0.8001 &0.8177 &0.8050 & 0.8198 &0.8040  &0.8163 \\
 
\bottomrule
\end{NiceTabular}}
\label{tab:generation_tasks}
\end{table*}
\subsection{Benchmarking on Literature Tasks}
\label{sec:benchmarkingsec}
%\xhdr{Evaluation Settings} In this subsection, we investigate the performance of various language models on elementary literature tasks constructed from LitBench corresponding to three domains: quantitative biology, robotics, and quantum physics. We consider a diverse range of open-source models, spanning from 1B to 8B parameters. We also consider SOTA closed-source models, specifically GPT-4o and DeepSeek-R1, to provide a comprehensive comparison of performance across both open and closed models. Additionally, we showcase the performance of a family of LLMs of different sizes ranging from 1B to 8B, trained using instructions datasets constructed from LitBench. The fine-tuning details are provided in Appendix \ref{appendix:trainingdetails}. 

\xhdr{Performance metrics}
To evaluate model performance across literature-related tasks, we employ different metrics tailored to each task type. For classification tasks, such as citation link prediction, we use standard accuracy metrics. For text generation tasks—such as citation sentence generation or abstract-to-title conversion—we selected BERTScore as our primary metric \cite{zhang2020bertscoreevaluatingtextgeneration}. Unlike BLEU \cite{10.3115/1073083.1073135} and ROUGE \cite{lin-2004-rouge}, which rely on exact matches to reference texts and fail to capture nuanced semantic differences, BERTScore captures intricate semantic differences without relying on strict adherence to ground truth text. Additionally, we excluded LLM-based evaluation, as it struggles to discern subtle differences between valid but domain-imperfect outputs, particularly when specific terms or structures are missing. Details on the inference pipeline for the performance evaluation are reported in Appendix \ref{appendix:trainingdetails}-2.
%For baseline evaluations, we consider a diverse range of open-source models, spanning from 1B to 8B parameters. This includes models from the LLaMA-3 family, as well as Vicuna-7B and Mistral-7B. Additionally, we evaluate SOTA closed-source models, specifically GPT-4 and DeepSeek-R1, to provide a comprehensive comparison of performance across both open and closed models.
%On the other hand, we fine-tune three models, Llama-3 family of 1, 3, and 8B, on each of the appropriate graphs corresponding to the domain. 

\begin{table}[t]
\centering
\caption{Evaluation of various models on predictive tasks across quantitative biology, robotics, and quantum physics.}
\resizebox{0.9\linewidth}{!}{
\begin{NiceTabular}{l|ccc|ccc}\toprule
 & \multicolumn{3}{c|}{\textbf{Citation Link Prediction}} & \multicolumn{3}{c}{\textbf{Citation Recommendation}} \\
 \midrule
 Datasets & \multicolumn{1}{c}{Biology} & \multicolumn{1}{c}{Robotics} & \multicolumn{1}{c|}{Quantum} &\multicolumn{1}{c}{Biology} & \multicolumn{1}{c}{Robotics} & \multicolumn{1}{c}{Quantum} \\ \midrule
Llama-3.2-1B & 16.17 & 16.77 & 22.32 & 4.40 & 1.96 & 1.60 \\
Llama-3.2-1B-Lit & 51.83 & 58.76 & 49.85 & 56.40 & 55.68 & 52.01 \\
\midrule
Llama 3.2-3B & 19.11 & 18.96 & 19.25 & 13.60 & 10.16 & 10.96 \\
Llama 3.2-3B-Lit & 79.22 & 73.43 & 70.05 & 72.42 & 63.90 & 64.70 \\
\midrule
Llama-3-8B & 36.53 & 36.41 & 36.74 & 19.23 & 17.75 & 19.51  \\
Llama-3-8B-Lit & 72.81 & 73.10 & 78.01 & 72.79 & 72.27 & 69.40 \\
\midrule
Mistral 7B & 50.90 & 51.32 & 50.66 & 27.94 & 18.80 & 15.50 \\
Vicuna 7B & 52.94 & 63.90 & 54.94 & 12.50 & 11.75 & 13.10 \\
\midrule
GPT-4o & 72.05 & 85.47 & 81.43 & 71.32 & 66.99 & 61.38 \\
DeepSeek-R1 & 68.30 & 70.05 & 59.35 & 73.30 & 93.30 & 84.30 \\
\bottomrule
\end{NiceTabular}}
\vspace{-15pt}
\label{tab:LP_tasks}
\end{table}

\begin{table*}[]
\caption{Comparison of specialized models and generically trained models across multiple domains.}
\centering
\resizebox{\linewidth}{!}{
\begin{NiceTabular}{c|c|c|c|cc|cc|cc|cc} 
\toprule 
\textbf{Datasets}
&\textbf{Models}
&\textbf{Citation Link Prediction}
&\textbf{Citation Recommendation}
&\multicolumn{2}{c|}{\textbf{Title Generation}}
&\multicolumn{2}{c|}{\textbf{Abstract Completion}}
&\multicolumn{2}{c|}{\makecell{\textbf{Citation Sentence} \\ \textbf{Generation}}}
&\multicolumn{2}{c}{\makecell{\textbf{Introduction} \\ \textbf{to Abstract}}} \\
\midrule
&& Accuracy & Accuracy & Precision & F1 Score & Precision & F1 Score & Precision & F1 Score & Precision & F1 Score \\
\midrule 
\multirow{2}{*}{Biology}
 &Llama-3.2-1B-Generic &48.27 &47.61 &0.7710  &0.8027  &0.8279 &0.8315 &0.8227 &0.8185  &0.8213 &0.8279 \\
 &Llama-3.2-1B-Lit  &51.83 &56.40 &0.7719 &0.8073  &0.8392 &0.8393  &0.8261 &0.8216 &0.8224  &0.8317  \\

\midrule
\multirow{2}{*}{Quantum}
 &Llama-3.2-1B-Generic &47.77 &44.44 &0.7848 &0.8179 &0.8275 &0.8373  &0.8225 &0.8153  &0.8180 &0.8259  \\
 &Llama-3.2-1B-Lit  &49.85 &52.01 &0.7860 &0.8189 &0.8412  &0.8455  &0.8291 &0.8212  &0.8300 &0.8343  \\

\midrule
\multirow{2}{*}{Robotics}
 &Llama-3.2-1B-Generic &51.98 &51.125 &0.7905 &0.8248  &0.8431 &0.8487 &0.8339  &0.8327  &0.8326 &0.8330  \\
 &Llama-3.2-1B-Lit  &58.76 &55.68 &0.8066 &0.8362 &0.8446 &0.8487  &0.8349 &0.8329  &0.8305 &0.8361  \\

\bottomrule
\end{NiceTabular}}
\label{tab:specialization_tasks}
\end{table*}
\begin{table*}[]

\vspace{-10pt}
\caption{Comparison of specialized models evaluated in the niche domain of AI applications in biology.}
\centering
\resizebox{\linewidth}{!}{
\begin{NiceTabular}{c|c|c|c|cc|cc|cc|cc} 
\toprule 
\textbf{Datasets}
&\textbf{Models}
&\textbf{Citation Link Prediction}
&\textbf{Citation Recommendation}
&\multicolumn{2}{c|}{\textbf{Title Generation}}
&\multicolumn{2}{c|}{\textbf{Abstract Completion}}
&\multicolumn{2}{c|}{\makecell{\textbf{Citation Sentence} \\ \textbf{Generation}}}
&\multicolumn{2}{c}{\makecell{\textbf{Introduction} \\ \textbf{to Abstract}}} \\
\midrule
&& Accuracy & Accuracy & Precision & F1 Score & Precision & F1 Score & Precision & F1 Score & Precision & F1 Score \\
\midrule 
\multirow{1}{*}{Artificial Intelligence}
 &Llama-3.2-1B-Lit  &51.47 &34.31 &0.7804 &0.8171 &0.8447 &0.8466  &0.8214 &0.8301 &0.8257  &0.8276  \\
\midrule
\multirow{1}{*}{Biology}
 &Llama-3.2-1B-Lit  &49.50 &36.01 &0.7765 &0.8102 &0.8412 &0.8461  &0.8277 &0.8285  &0.8264 &0.8259  \\
\midrule
\multirow{1}{*}{Application of AI in Biology}
 &Llama-3.2-1B-Lit  &55.39 &39.29 &0.7855 &0.8208 &0.8474 &0.8495  &0.8319 &0.8336  &0.8294 &0.8310  \\

\bottomrule
\end{NiceTabular}}
\label{tab:extreme_specialization}
\end{table*}

\begin{comment}
\begin{table*}[]
\caption{}
\centering
\resizebox{\linewidth}{!}{
\begin{NiceTabular}{c|c|c|c|c|c|c|c} 
\toprule 
\textbf{Datasets}
&\textbf{Models}
&\multicolumn{1}{c}{\textbf{Citation Link Prediction}}
&\multicolumn{1}{c}{\textbf{Citation Recommendation}}
&\multicolumn{1}{c}{\textbf{Title Generation}}
&\multicolumn{1}{c}{\textbf{Abstract Completion}}
&\multicolumn{1}{c}{\textbf{Citation Sentence Generation}}
&\multicolumn{1}{c}{\textbf{Introduction to Abstract}}
\\
\midrule 
\multirow{2}{*}{Biology}
 &Llama-3.2-1B-Ablation &52.25 &58.85  &0.8014 &0.8301 &0.8168  &0.8286\\
 &Llama-3.2-1B-Lit  &51.83 &56.4  &0.8073 &0.8393 &0.8216  &0.8317\\

\midrule
\multirow{2}{*}{Quantum}
 &Llama-3.2-1B-Ablation &49.61 &53.35 &0.8012 &0.8415 &0.8188  &0.8318\\
 &Llama-3.2-1B-Normal  &49.85 &52.01  &0.8189 &0.8455 &0.8212  &0.8343\\

\midrule
\multirow{2}{*}{Robotics}
 &Llama-3.2-1B-Ablation &58.55 &57.53  &0.8158 &0.8373 &0.8252  &0.8338\\
 &Llama-3.2-1B-Lit  &58.76 &55.68 &0.8362 &0.8487 &0.8329  &0.8361\\

\bottomrule
\end{NiceTabular}}
\label{tab:ablation_tasks}
\end{table*}
\end{comment}

\begin{table*}[]
\vspace{-10pt}
\caption{Performance comparison of individually tuned models vs. unified instruction tuning}
\centering
\resizebox{\linewidth}{!}{
\begin{NiceTabular}{c|c|c|c|cc|cc|cc|cc} 
\toprule 
\textbf{Datasets}
&\textbf{Models}
&\textbf{Citation Link Prediction}
&\textbf{Citation Recommendation}
&\multicolumn{2}{c|}{\textbf{Title Generation}}
&\multicolumn{2}{c|}{\textbf{Abstract Completion}}
&\multicolumn{2}{c|}{\makecell{\textbf{Citation Sentence} \\ \textbf{Generation}}}
&\multicolumn{2}{c}{\makecell{\textbf{Introduction} \\ \textbf{to Abstract}}} \\
\midrule
&& Accuracy & Accuracy & Precision & F1 Score & Precision & F1 Score & Precision & F1 Score & Precision & F1 Score \\
\midrule 
\multirow{2}{*}{Biology}
 &Llama-3.2-1B-Indiv &52.25 &58.85 &0.7841  &0.8014  &0.8200 &0.8301 &0.8230 &0.8168  &0.8256 &0.8286  \\
 &Llama-3.2-1B-Lit  &51.83 &56.40 &0.7719 &0.8073  &0.8392 &0.8393  &0.8261 &0.8216 &0.8224  &0.8317  \\

\midrule
\multirow{2}{*}{Quantum}
 &Llama-3.2-1B-Indiv &49.61 &53.35 &0.7705 &0.8012 &0.8255 &0.8415  &0.8166 &0.8188  &0.8310 &0.8318  \\
 &Llama-3.2-1B-Lit  &49.85 &52.01 &0.7860 &0.8189 &0.8412  &0.8455  &0.8291 &0.8212  &0.8300 &0.8343  \\

\midrule
\multirow{2}{*}{Robotics}
 &Llama-3.2-1B-Indiv &58.55 &57.53 &0.7811 &0.8158  &0.8394 &0.8373 &0.8306  &0.8252  &0.8322 &0.8338  \\
 &Llama-3.2-1B-Lit  &58.76 &55.68 &0.8066 &0.8362 &0.8446 &0.8487  &0.8349 &0.8329  &0.8305 &0.8361  \\

\bottomrule
\end{NiceTabular}}
\label{tab:ablation_tasks}
\end{table*}

\xhdr{Performance on literature tasks} 
As shown in Table \ref{tab:generation_tasks} and \ref{tab:LP_tasks}, fine-tuning LLMs on LitBench significantly enhances their performance across the various literature tasks. This improvement is most pronounced for smaller models, though it remains substantial even for larger ones. Particularly, fine-tuned models can outperform larger models like GPT-4o and DeepSeek-R1 on many of these tasks.  This is attributed to LitBench's ability to allow the models to internalize intricate domain-specific knowledge and structural patterns found in the graph, thus enabling them to generalize effectively even to test nodes not encountered during training.

Another key observation from these tables is the consistency of these results across all domains. However, the extent of performance gains depends on the degree of domain-specific information internalized during pre-training by these models, with greater improvements observed in domains where prior internalization was limited. %Finally, while models like GPT-4o may retain an advantage in English language formulation—potentially outperforming fine-tuned models on certain tasks when evaluated using BERTScore—fine-tuned models consistently excel in tasks where domain-specific graph structure and knowledge are critical, such as link prediction and node retrieval tasks.
Finally, while models like GPT-4o may retain an advantage in English language formulation and DeepSeek-R1 may excel in tasks such as citation recommendation due to its test-time scaling capabilities—allowing them to potentially outperform fine-tuned models on certain tasks—fine-tuned models consistently excel in literature tasks where domain-specific graph structure and knowledge are critical. Moreover, these fine-tuned models achieve competitive performance at a fraction of the size of their larger counterparts, demonstrating the effectiveness of domain-specific adaptation.

%the general trend concerns increasing performance across these tasks increases consistently as the size of the model goes up. Barring a few minor violations to this general rule, and minor fluctuations, this is the first conclusion that is consistent of what one would expect. 
%The second thing that can be concluded from these results is 

\xhdr{Performance on advanced downstream tasks} Beyond the above tasks, we investigate how LitBench enables tackling more advanced ones, such as related work generation. The complexity of these tasks presents significant evaluation difficulties for previously used metrics. For instance, evaluating the generation of related work involves assessing citation diversity and the relevance of cited material—both of which are challenging to quantify with single metrics. To address this, we provide case studies along with a structured human evaluation framework to assess model performance comprehensively.

For the related work generation task, we compare the strongest baseline models (GPT-4o and DeepSeek-R1) against an 8B-parameter Llama model fine-tuned on LitBench across robotics, quantum physics, and quantitative biology domains. The fine-tuned model operates in a chain-of-thought fashion, with retrieval access to the LitBench dataset to generate responses. Our findings demonstrate that the LitBench-fine-tuned model outperforms GPT-4o and DeepSeek-R1 in terms of topic diversity, depth of detail, and the diversity of cited material. However, it lags behind in English structure and coherence, which is expected given the significant size discrepancy between the models. The detailed evaluation methodology is provided in Appendix \ref{appendix:humaneval} and the case studies are provided in Appendix \ref{appendix:casestudies}.

Moving to the task of identifying influential papers, LitBench's concept modeling enables us to identify relevant subgraphs and reason over them effectively. Specifically, we can pinpoint the most influential papers in a domain by retrieving those with the highest in-degree (i.e., the most cited papers). In contrast, when prompted to identify influential papers, LLMs often generate fabricated or non-influential references. Case studies illustrating these issues are provided in Appendix \ref{appendix:casestudy}. 

\subsection{LitBench for Specialization}

\xhdr{The Need for Specialization}
LitBench is designed to create domain-specific subgraphs that not only enable the evaluation of models on literature tasks within a particular domain but also facilitate the generation of training datasets for developing domain-specific literature models. A natural question arises: is such specialization necessary compared to a generic approach, where models are trained on random pairs of literature tasks regardless of domain? To address this, we benchmark the performance of two LLaMA-3.2-1B models: one fine-tuned on domain-specific subgraphs (as described in Section \ref{sec:benchmarkingsec}) and another trained on random pairs of literature tasks from unrelated domains. Both models are trained under identical conditions, including dataset size and hyperparameters found in Appendix \ref{appendix:trainingdetails}-1. 

As shown in Table \ref{tab:specialization_tasks}, the domain-specific model consistently outperforms the generic model across all tasks. While training on random papers does improve performance by allowing the model to better leverage its pretrained knowledge and extract structural patterns, it still falls short of the domain-specific approach. This underscores the importance of domain specialization in achieving superior performance on literature-related tasks.

\xhdr{Pushing the limits of Specialization}
One key feature of LitBench is its ability to create domain-specific subgraphs for any field, no matter how niche. Beyond high-level domains like robotics, biology, and quantum physics, LitBench can be tailored to highly specialized areas, enabling the creation of literature experts 

To demonstrate this, we consider the niche domain of "AI applications in biology." Using LitBench, we curate a subgraph for this specific area and use it to generate instructions for training a LLaMA-3.2-1B model, with the same training settings as in the previous subsection. For comparison, we also train models on subgraphs related to the broader domains of ``Biology'' and ``Artificial Intelligence,'' with each subgraph containing 10,000 retrieved nodes.

As shown in Table \ref{tab:extreme_specialization}, the model trained on the niche domain of ``AI applications in biology'' outperforms the models trained on the broader domains across all tasks. This highlights the power of LitBench to push the boundaries of specialization, enabling better performance even in highly niche areas of interest.
\subsection{Ablation Studies}
\label{sec:ablation}
\begin{table}[t]
\centering
\caption{Comparison of retrieval performance between titles and abstracts vs. LitBench's concepts field.}
\resizebox{0.9\linewidth}{!}{
\begin{NiceTabular}{l|ccc|ccc}\toprule
 & \multicolumn{3}{c|}{\textbf{Concepts}} & \multicolumn{3}{c}{\textbf{Title + Abstract}} \\
 \midrule
 Datasets & \multicolumn{1}{c}{Biology} & \multicolumn{1}{c}{Robotics} & \multicolumn{1}{c|}{Quantum} &\multicolumn{1}{c}{Biology} & \multicolumn{1}{c}{Robotics} & \multicolumn{1}{c}{Quantum} \\ \midrule
BAAI & 74.87 & 92.21& 95.05 & 8.7 & 3.45 & 5.7 \\
\bottomrule
\end{NiceTabular}}
\vspace{-15pt}
\label{tab:retrieval task}
\end{table}

\begin{table*}[]
\caption{Performance comparison of finetuned and instruction tuned models versus instruction tuned model across quantitative biology, robotics, and quantum physics domains.}
\centering
\resizebox{\linewidth}{!}{
\begin{NiceTabular}{c|c|c|c|cc|cc|cc|cc} 
\toprule 
\textbf{Datasets}
&\textbf{Models}
&\textbf{Citation Link Prediction}
&\textbf{Citation Recommendation}
&\multicolumn{2}{c|}{\textbf{Title Generation}}
&\multicolumn{2}{c|}{\textbf{Abstract Completion}}
&\multicolumn{2}{c|}{\makecell{\textbf{Citation Sentence} \\ \textbf{Generation}}}
&\multicolumn{2}{c}{\makecell{\textbf{Introduction} \\ \textbf{to Abstract}}} \\
\midrule
&& Accuracy & Accuracy & Precision & F1 Score & Precision & F1 Score & Precision & F1 Score & Precision & F1 Score \\
\midrule 
\multirow{2}{*}{Biology}
 &Llama-3.2-1B-Lit  &51.83 &56.40 &0.7719 &0.8073  &0.8392 &0.8393  &0.8261 &0.8216 &0.8224  &0.8317  \\
  &Llama-3.2-1B-FT-Lit &53.67 &62.50 &0.7867  &0.8121  &0.8415 &0.8410 &0.8268 &0.8234  &0.8223 &0.8318  \\

\midrule
\multirow{2}{*}{Quantum}
 &Llama-3.2-1B-Lit  &49.85 &52.01 &0.7860 &0.8189 &0.8412  &0.8455  &0.8291 &0.8212  &0.8300 &0.8343  \\
  &Llama-3.2-1B-FT-Lit &50.80 &53.16 &0.7943 &0.8227 &0.8437 &0.8455  &0.8255 &0.8176  &0.8290 &0.8344  \\

\midrule
\multirow{2}{*}{Robotics}
 &Llama-3.2-1B-Lit  &58.76 &55.68 &0.8066 &0.8362 &0.8446 &0.8487  &0.8349 &0.8329  &0.8305 &0.8361  \\
  &Llama-3.2-1B-FT-Lit &59.33 &56.22 &0.7994 &0.8292  &0.8473 &0.8506 &0.8353  &0.8332 &0.8294 &0.8350  \\

\bottomrule
\end{NiceTabular}}
\vspace{-10pt}
\label{tab:pretraining}
\end{table*}
\xhdr{LitBench's Concepts for Retrieval}
In our first ablation study, we demonstrate the usefulness of LitBench's concepts field in retrieving domain-specific subgraphs relevant to a user's query. This effectiveness is evaluated across three distinct user queries requesting subgraphs in the following domains: quantitative biology, quantum physics, and robotics. To establish our ground truth dataset, we sampled 30,000 random papers from the relevant categories on arXiv. For embedding, we employ BGE-large, a state-of-the-art encoder model optimized for retrieval tasks.
%\footnote{\url{https://arxiv.org/category_taxonomy}}

Our evaluation compared two approaches: a baseline method using combined title and abstract embeddings, and our proposed method using LitBench's concepts field embeddings. Using cosine similarity, we retrieve the top 30,000 papers using each method and measure recall for both methods compared to the ground truth data of the corresponding domain. As shown in Table \ref{tab:retrieval task}, our approach significantly outperforms the baseline. In fact, titles and abstracts often suffer from information overload, which can degrade embedding quality. In contrast, LitBench's concepts field mitigates this by mapping each paper to a concise list of aligned concepts, providing a cleaner and more effective representation for retrieval. For this reason, LitBench's concepts field enables the creation of accurate domain-specific subgraphs relevant to the user's desired field.

\xhdr{Transfer Learning Capabilities} In our second ablation study, we investigate the transfer learning capabilities of the multi-instruction approach. Specifically, LitBench offers two utilization strategies: training on individual tasks to address specific user needs or training within a unified multi-instruction framework. We argue that the unified approach enables models to internalize both the attributes of the graph and its structural relationships through link information and their associated attributes, thereby enhancing generalization and transfer learning potential. Therefore, a key question is how effective this unified approach is in improving performance on literature tasks compared to training on individual tasks in isolation. To address this, we evaluate LLaMA-3.2-1B using the same training settings as in the previous subsection. We compare the performance of a model trained on the entire LitBench framework against models trained separately on individual tasks (referred to as the -Indiv model), thus assessing the benefits of the unified approach in terms of task performance and domain adaptability.

As shown in Table \ref{tab:ablation_tasks}, the unified framework demonstrates clear advantages, particularly in generative tasks such as abstract completion and abstract-to-introduction generation, where domain-specific knowledge proves highly beneficial. However, this transfer learning effect is less pronounced in predictive tasks like link prediction, where task-specific training remains competitive. Therefore, it remains best practice to utilize the unified framework to extract the most out of the LitBench dataset.

\xhdr{Pretraining as Precursor} 
By design, our proposed tool allows users to save the full cleaned content of a paper as a node attribute in the created domain subgraph, if desired. This raises another critical question: is continual pretraining on the domain of interest necessary to further improve performance? Prior work has demonstrated that pretraining on a target domain before task-specific adaptation can enhance model performance \cite{gururangan-etal-2020-dont}. To test this hypothesis, we benchmark the approach using LLaMA-3.2-1B on the three curated domains of quantum physics, quantitative biology, and robotics. In this setup, pretraining is performed on the full text of papers to familiarize the model with the target domain, followed by fine-tuning on instructions derived from LitBench. The pretraining details are provided in Appendix \ref{appendix:trainingdetails}-3.

As shown in Table \ref{tab:pretraining}, pretraining on unstructured domain material prior to fine-tuning on LitBench does yield a performance improvement. However, the gains are minimal. Given the substantial computational resources required for continual pretraining, we conclude that this additional step is not necessary for achieving strong performance on domain-specific literature tasks. Training on instructions from LitBench alone is sufficient to deliver robust and competitive results.

\xhdr{LitBench Graph Size}
When constructing LitBench, an important parameter to determine is the number of retrieved nodes required to build an effective dataset. To address this, we evaluate three subgraphs representing the domains of quantitative biology, quantum physics, and robotics, training both LLaMA-3.2-1B and LLaMA-3.2-3B models. We maintain the same training settings described in the previous section. 

Our results showcase that these models learn quickly, with performance plateauing after approximately 1000 training steps, with larger models requiring fewer steps to converge. Given our graph sizes, instruction set, and batch size, these 1,000 steps corresponded to traversing approximately 1,000 nodes. While the exact number of required nodes to be retrieved may vary by domain, the consistent finding is that only a small subset of papers from the domain is sufficient to internalize the domain-specific knowledge found in their graph. It is worth noting that our tool allows the user to select the desired number of retrieved nodes. For reference, the training losses are reported in Fig. \ref{fig:graphsize} found in Appendix \ref{appendix:traininggraphsize}.

\section{Conclusion}
In this work, we introduced LitBench, a benchmarking tool designed to enable the development and evaluation of domain-specific LLMs tailored to literature-related tasks. LitBench emphasizes the importance of domain-specific adaptation to achieve strong performance on literature graphs. Our evaluation demonstrated that LitBench is highly flexible, allowing the creation of training and evaluation datasets for any user-specified topic, no matter how niche. This capability enables the development of specialized LLM literature agents that rival state-of-the-art models in performance, even at a fraction of their size.

% \section{Acknowledgements}
\begin{acks}
This research was supported in part by the National Science Foundation (NSF) Division of Computer and Network Systems (CNS-2431504 and CNS-2402862), NSF FAIN 2132573, and the Army Research Office under Contract W911NF-23-1-0088. We also acknowledge support in part from the Yale AI Engineering Research Grant from the Yale Office of the Provost, the Yale Engineering AI Seed Grant, the Roberts Innovation Award (2025), sponsored research from Goldman Sachs, and the AWS Research Awards.
\end{acks}

\bibliographystyle{ACM-Reference-Format}
\bibliography{references}

\section*{Appendices}
\appendix

\section{State-of-the-Art Datasets}
\label{appendix:datasetssota}
Several datasets have been proposed to support machine learning on scientific literature. While they offer useful features, none include the full set of fields necessary for training and evaluating LLMs across the range of citation prediction and generation tasks covered in LitBench. These missing elements directly impact the model’s ability to recommend relevant papers, predict citations, and transfer knowledge across these tasks. We provide below an in-depth discussion of each of these datasets:

   \noindent\textbf{1. MAG} was introduced in \citep{sinha2015overview} as a large-scale academic knowledge graph, providing structured metadata such as titles, abstracts, and citation links. It has been widely used in graph-based approaches for paper ranking and influence estimation \citep{10.1145/3308558.3313700}. However, it lacks several elements crucial for citation reasoning tasks. Firstly, it does not include citation sentences, so models cannot learn how citations are expressed in context. Also, there is no section-level structure (e.g., introductions or related work), making it unsuitable for training models on abstract completion or introduction-to-abstract generation. Finally, concept annotations are missing, which prevents the retriever from ranking candidate papers by topical relevance; this leads to mostly irrelevant negative samples in citation link prediction and degrades performance on citation recommendation.
   
    \noindent\textbf{2. ArXiv} dataset \citep{saier2023unarxive} was constructed by crawling raw LaTeX files or PDFs from ArXiv. They offer access to full paper content, but without citation alignment, section tagging, or concepts. To make them usable for structured tasks, one must parse the documents to extract citation edges, locate sections, and match references to in-text citations—an error-prone and time-consuming process. Without a consistent structure, models cannot transfer knowledge between sections, which is necessary for abstract completion or introduction-to-abstract generation.

    \noindent\textbf{3. OpenAlex} was proposed in \citep{priem2022openalex} as a successor to MAG, improving metadata coverage and adding high-level concept tags to papers. It has been useful for large-scale bibliometric analysis and concept-based exploration \citep{10.1371/journal.pone.0308041}. However, it lacks in-text citation sentences and any section-level content, which limits its utility for training LLMs on tasks like citation sentence generation or abstract completion.

    \noindent\textbf{4. S2ORC} \cite{lo2019s2orc} is one of the most text-rich literature datasets, containing full papers, section labels, and aligned citation sentences. It has been used in a number of citation-aware modeling tasks \citep{roy2024ilciterevidencegroundedinterpretablelocal}. However, it does not include concept annotations, making it difficult to construct domain-specific subgraphs or perform concept-aware filtering. This affects both retrieval and training: in citation link prediction, negative samples are typically drawn from clearly unrelated domains, reducing the challenge of the task and the ability of the model to extract intricate relationships among papers. In contrast, LitBench uses concepts, thus sampling harder negatives—similar yet unrelated  papers—which leads to better model discrimination. Furthermore, the lack of consistent related work and introduction sections limits potential transfer learning across the literature tasks.

With that in mind, LitBench is the only tool that creates literature dataset that brings together all necessary components: titles, abstracts, introductions, related work, citation sentences, and concept annotations. This complete structure enables accurate citation modeling, hard negative sampling, and transfer learning between literature tasks for any user-defined domain, no matter how niche.

% This far ... 

% Go into details about what each one lacks and how it differs from our dataset and what makes our dataset good as we allow transfer learning bla bla. Let us be specific about what makes it different and how it will translate it in better performance across these different domains. 

% Bullet points:...

% M2G
% S2ORC lacks concepts, makes it difficult. By taking samples here, the negative samples are easier compared to harder sampling. 

\section{Concepts Generation Prompts}
\label{appendix:prompts}
The prompts below were employed to generate LitBench concept fields, spanning three abstraction levels (high-level disciplines, mid-level scientific fields, and low-level thematic topics):
\setlistingname{Prompt}
\raggedbottom
\begin{lstlisting}[style=promptstyle, caption={Abstract Level 1}, basicstyle=\ttfamily] 
Given a scientific paper, identify the three major academic disciplines or broad scientific fields that are most relevant to the paper's content. Avoid sub-disciplines or specific research areas. Focus on the fundamental scientific domains that encompass the paper's core concepts and methodologies. Present these disciplines in a simple numbered list. Here is the title and abstract of the paper: {Title}{Abstract}
\end{lstlisting}
\raggedbottom
\begin{lstlisting}[style=promptstyle, caption={Abstract Level 2}, basicstyle=\ttfamily]
Given a scientific paper, identify the three major scientific fields that are most relevant to the paper's content. Present these disciplines in a simple numbered list. Here is the title and abstract of the paper: {Title}{Abstract}
\end{lstlisting}
\raggedbottom
\begin{lstlisting}[style=promptstyle, caption={Abstract Level 3}, basicstyle=\ttfamily]
Given a scientific paper, generate a list of three very high-level topics of maximum three words that summarize the main areas covered in the paper. These topics should be broad categories that capture the key themes of the paper. Present these disciplines in a simple numbered list. Here is the title and abstract of the paper: {Title}{Abstract}
\end{lstlisting}

\section{Concepts Examples}
\label{appendix:concepts}
Below, we provide a set of examples illustrating how concepts are assigned to each paper across different abstraction levels.

\begin{tcolorbox}[colframe=niceblue!75!black, colback=blue!5!white, sharp corners=south, title=Example 1]
\textbf{Title:} A quasi classical approach to electron impact ionization

\textbf{Abstract:} A quasi classical approximation to quantum mechanical scattering in the Moeller formalism is developed. While keeping the numerical advantage of a standard Classical--Trajectory--Monte--Carlo calculation, our approach is no longer restricted to use stationary initial distributions. This allows one to improve the results by using better suited initial phase space distributions than the microcanonical one and to gain insight into the collision mechanism by studying the influence of different initial distributions on the cross section. A comprehensive account of results for single, double and triple differential cross sections for atomic hydrogen will be given, in comparison with experiment and other theories.

\textbf{Level 1:} Physics, Mathematics, Chemistry

\textbf{Level 2:}  Quantum Mechanics, Theoretical Physics,
            Atomic Physics

\textbf{Level 3:}  Quantum Scattering Theory, Classical Trajectory Methods, Atomic Collision Dynamics
\end{tcolorbox}

\begin{tcolorbox}[colframe=niceblue!75!black, colback=blue!5!white, sharp corners=south, title=Example 2]
\textbf{Title:} Beyond mean-field boson-fermion description of odd nuclei

\textbf{Abstract:} We develop a novel theoretical method for calculating spectroscopic properties of those nuclei with odd number of nucleons, that is based on the nuclear density functional theory and the particle-boson coupling scheme. Self-consistent mean-field calculation based on the DFT is performed to provide microscopic inputs to build the Hamiltonian of the interacting boson-fermion systems, which gives excitation spectra and transition rates of odd-mass nuclei. The method is successfully applied to identify the quantum shape phase transitions and the role of octupole correlations in odd-mass nuclei, and is extended further to odd-odd nuclear systems.

\textbf{Level 1:} Physics, Mathematics, Chemistry

\textbf{Level 2:}  Nuclear Physics, Theoretical Physics, Quantum Mechanics

\textbf{Level 3:}  Nuclear Density Functional, Particle-Boson Coupling, Nuclear Spectroscopy
\end{tcolorbox}

\section{Raw Data Processing}
\label{sec:appendix_cleaning}

After identifying relevant arXiv papers, the download and processing phase begins. The process involves downloading each paper and removing all comments from the LaTeX files using Google's arXiv LaTeX Cleaner\footnote{\url{https://github.com/google-research/arxiv-latex-cleaner}}.

Next, since LaTeX sources often consist of multiple files, we unify the LaTeX commands used for importing these files, ensuring that all imports use the \textbackslash input\{$\cdot$\} command. By constructing a directed graph to map these relationships, we identify the main LaTeX file of the paper. We then flatten the document using the Latexpand Perl script\footnote{We tested fatex (\url{https://ctan.org/pkg/fatex}) and fap (\url{https://github.com/fchauvel/fap}), but achieved the best results with latexpand (\url{https://ctan.org/pkg/latexpand}).}, consolidating all content into the main file.

In the subsequent step, we address custom commands defined by authors through a 'de-macoring' process. We unify all custom commands—such as \textbackslash def\{$\cdot$\} and \textbackslash DeclareMathOperator\{$\cdot$\}—and replace them \textbackslash newcommand\{$\cdot$\} before substituting them with their native LaTeX equivalents using the Python library de-macro\footnote{\url{https://ctan.org/pkg/de-macro}}. We then remove figures, tables, and over a hundred non-informative LaTeX commands and environments using regular expression matching, ensuring the final LaTeX files contain only text and equations without unnecessary LaTeX residue. After completing this preprocessing step, we identify the introduction and related work sections in each paper and create corresponding node attributes for these sections in the resulting sub-graph.

Additionally, we standardize all citation commands, extract citation labels, and map them to the .bib file in the LaTeX project to obtain citation metadata, particularly titles. If no .bib file is available, we search for .bbl files and fall back on in-line citations using the LaTeX `thebibliography' environment if necessary. The extracted metadata allows us to identify the cited materials in the arXiv graph, thus forming citation links in the created sub-graph. Once a link is detected, we also extract the citation sentence and add it to the citation edge as a textual attribute.

By applying this process across all retrieved papers, we construct our desired domain-specific subgraph.

\section{Datasets Preparation}
\label{appendix:datasetprep}
Given a domain-specific subgraph in LitBench, we construct training samples as follows. We create instruction datasets for each task by leveraging node and edge attributes. Specifically, we use the title and abstract of each paper to generate instructions for title generation, abstract completion, and introduction-to-abstract tasks.

For edge-related tasks, each edge in LitBench is used to create a positive citation link prediction instruction. To balance the dataset, we replace the target node of each edge with a random node to construct a corresponding negative citation link prediction instruction, maintaining a 1:1 ratio of positive to negative samples. Additionally, for each citation sentence associated with an edge, we construct an instruction that takes the titles and abstracts of both nodes connected by the edge and generates the citing sentence.

Finally, for citation recommendation tasks, we randomly sample 10 negative nodes for each edge to form a candidate set, requiring the model to identify the positive candidate from the set.

\begin{figure*}[!t]
    \centering
    \includegraphics[width=0.86\textwidth]{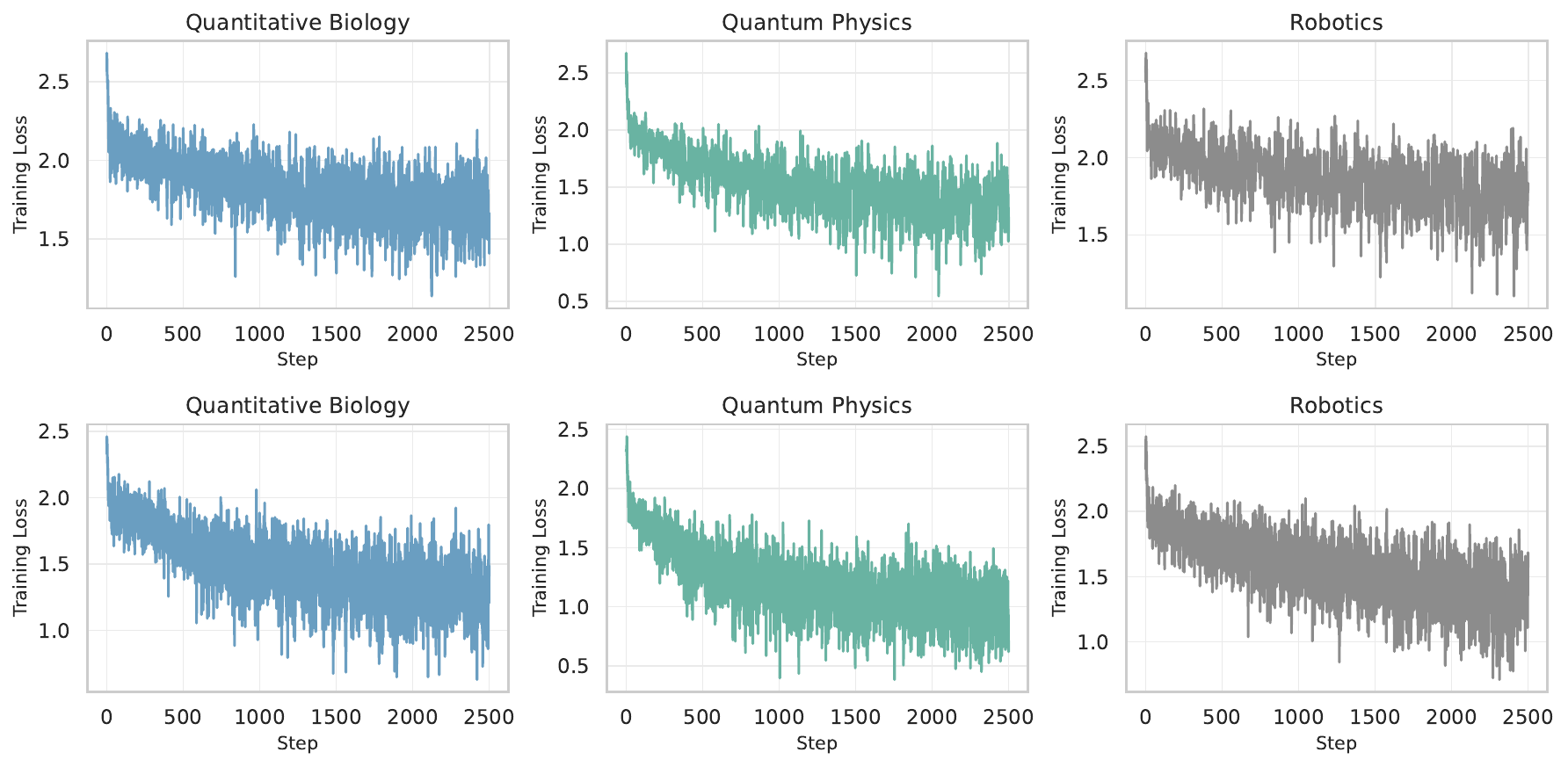}
    \caption{Training loss for LLama-3.2-1B (top) and LLama-3.2-3B (bottom) models across quantitative biology, robotics, and quantum physics domains.
    }
    \label{fig:graphsize}
\end{figure*}

\section{Hyperparameters}
\label{appendix:trainingdetails}
\xhdr{1. Fine-tuning settings} During fine-tuning, we employ the Q-LoRa approach 
\begin{comment}
\cite{dettmers2023qloraefficientfinetuningquantized} 
\end{comment}
where the model was loaded in 8-bit precision and the rank of LoRa was set to 8. The LoRa layers were applied exclusively to the $\mathbf{Q}$, $\mathbf{K}$, $\mathbf{V}$, and $\mathbf{O}$ matrices. We configured the batch size to 8 and used a gradient accumulation step of 2. The learning rate was set to 0.0002, with AdamW 
\begin{comment}
\cite{loshchilov2019decoupledweightdecayregularization}
\end{comment}
chosen as the optimizer. Additionally, we set the LoRa scaling hyperparameter $lora\_alpha$ to 32, and the dropout rate during fine-tuning is set to 0.05. 

\xhdr{2. Pipeline parameters} During the evaluation of models on literature tasks, we employed greedy decoding. For link prediction tasks, the number of new tokens generated was limited to 2. For title generation, the limit was set to 100 tokens, while abstract completion and introduction-to-abstract generation were capped at 256 tokens each. Citation sentence generation was restricted to 64 tokens. These token limits were chosen to balance task requirements and computational efficiency.

\xhdr{3. Pretraining settings} We configure the batch size to 4 million tokens, with each sequence containing 8,192 tokens. The optimization is handled by AdamW with a weight decay of 0.1, while retaining Hugging Face’s default $\beta$ parameters. Gradient clipping is applied with a maximum norm of 1. The learning rate follows a cosine decay schedule, gradually reducing to 10\% of its peak value, with a linear warmup phase spanning 10\% of an epoch. 
\begin{comment}
    \cite{gupta2023continual}
\end{comment}
The peak learning rate is set to 1e-5.

\begin{comment}
    
\section{Dataset Statistics}
\label{appendix:datasets}

Since , the number of nodes in the final graph , and some papers cannot be downloaded due to being either withdraw, or LaTex sources not available, we only keep connected nodes in the graph, the number

Based on these distributions, we select 30,000 papers per domain to construct domain-specific subgraphs.

We then generate literature tasks for both fine-tuning and evaluation following the methodology described earlier. Table I summarizes the training token count for each domain.

\end{comment}

\section{Human Evaluation}
\label{appendix:humaneval}

We designed a comparative assessment using Qualtrics, engaging AI researchers to evaluate related work sections generated by LLama-3-8B-Lit, GPT-4o, and DeepSeek-R1. Participants reviewed outputs for ten distinct papers across all considered domains and rated each model on six criteria using a relative scoring system (1–5), where higher scores indicated stronger performance. The evaluation metrics included:\\
\textbf{Q1:} Alignment with the paper’s title and abstract\\
\textbf{Q2:} Cited material diversity\\
\textbf{Q3:} Clarity, structure, and readability\\
\textbf{Q4:} Level of detail of the section \\
\textbf{Q5:} Appropriateness of the citations number

Researchers were instructed to assign scores comparatively, allowing tied ratings for comparable outputs. Results, summarized in Table \ref{tab:survey}, reveal LLama-3-8B-Lit's superiority in citation diversity  and depth of analysis. This model however, lagged in linguistic coherence, outperformed by GPT-4o and DeepSeek-R1, due to the latter models’ larger parameter scales enhancing language fluency.

\begin{table}[h!]
\caption{Human evaluation on related work generation task.}
\centering
\resizebox{0.7\linewidth}{!}{
\begin{NiceTabular}{|c|ccccc|} 
\toprule 
\textbf{Models}
&\multicolumn{1}{c}{Q1}
&\multicolumn{1}{c}{Q2}
&\multicolumn{1}{c}{Q3}
&\multicolumn{1}{c}{Q4}
&\multicolumn{1}{c}{Q5}\\
\midrule 
DeepSeek-R1  & \textbf{5} &	4.2	&4.6 &	3.6&	3.6	  \\
GPT-4o  &\textbf{5}	&4.3	&\textbf{5}	&3.8	&3.8	\\
 \midrule
 \midrule
Llama-3-8B-Lit & \textbf{5} &	\textbf{4.8} &	4.2	&\textbf{4.6}&	\textbf{4.8}	 \\
\bottomrule
\end{NiceTabular}}
\label{tab:survey}\vspace{-0.3cm}
\end{table}

\section{Influential Papers Case Studies}
\label{appendix:casestudy}

These case studies highlight how even SOTA LLMs struggle to accurately identify influential papers in niche domains, often succumbing to hallucinations. This underscores the utility of LitBench, which constructs domain-specific subgraphs capable of addressing such tasks through straightforward reasoning over the graph structure. Below, we present two examples of hallucinations from GPT-4o and DeepSeek-R1 when prompted about identifying influential papers involving applications of AI in biology. 
\vspace{-0.11cm}
\begin{tcolorbox}[
    colframe=customteal!75!black,  % Border color (Teal Green mixed with black)
    colback=custompeach!10!white,  % Background color (Peach mixed with white)
    sharp corners=south, 
    title=Example 1
]
\textbf{Model:} GPT-4o

\textbf{Query:} What are the most influential papers on applications of AI in biology?

\textbf{Answer:} AI has made significant contributions to biology, particularly in genomics, protein structure prediction, drug discovery, and medical diagnostics. Here are some of the most influential papers in AI applications to biology:
$$\dots$$
2. Genomics \& DNA Sequence Analysis \\ 
“Attention is all you need for genomics: Using transformers for large-scale gene sequence modeling” (Nature Communications, 2021) $\text{{\color{red}!! Hallucinated Citation!!}}$
$$\dots$$
\end{tcolorbox}

\begin{tcolorbox}[
    colframe=customteal!75!black,  % Border color (Teal Green mixed with black)
    colback=custompeach!10!white,  % Background color (Peach mixed with white)
    sharp corners=south, 
    title=Example 2
]
\textbf{Model:} DeepSeek-R1

\textbf{Query:} What are the most influential papers on applications of AI in biology?

\textbf{Answer:} The application of AI in biology has led to groundbreaking advancements in areas such as genomics, drug discovery, protein folding, and personalized medicine. Below are some of the most influential papers that have shaped the field:
$$\dots$$
4. Single-Cell Analysis \\
Single-Cell RNA Sequencing Data Analysis with Deep Learning \\
Authors: Lopez et al. (2018) \\
Journal: Nature Methods $\text{{\color{red}!! Hallucinated Citation!!}}$
$$\dots$$
\end{tcolorbox}

\section{Literature Graph Size}
\label{appendix:traininggraphsize}
In this appendix, we report the training loss curves in Fig. \ref{fig:graphsize} from our ablation study investigating the impact of literature graph size, as discussed in Section \ref{sec:ablation}.

\raggedbottom

\newpage

\onecolumn
\section{Related Work Generation Case Studies}
\label{appendix:casestudies}

% Switch to one-column mode if using two-column format

%%%%%% ROBOTICS %%%%%%%%%
{\renewcommand{\arraystretch}{1.3}

\begin{longtable}{|>{\centering\arraybackslash}m{2cm}|>{\centering\arraybackslash}m{14cm}|} 
\hline
\textbf{Model} & \textbf{Generated Related Work} \\
\hline
\endfirsthead

% Header for continued pages
\multicolumn{2}{c}{{\textit{(Continued from previous page)}}} \\
\hline
\textbf{Model} & \textbf{Generated Related Work} \\
\hline
\endhead

% Footer for all but last page
\hline \multicolumn{2}{|r|}{{Continued on next page}} \\ \hline
\endfoot

% Footer for last page
\hline
\endlastfoot

\textbf{Llama-3-8B-Lit} & \includegraphics[width=1.01\linewidth]{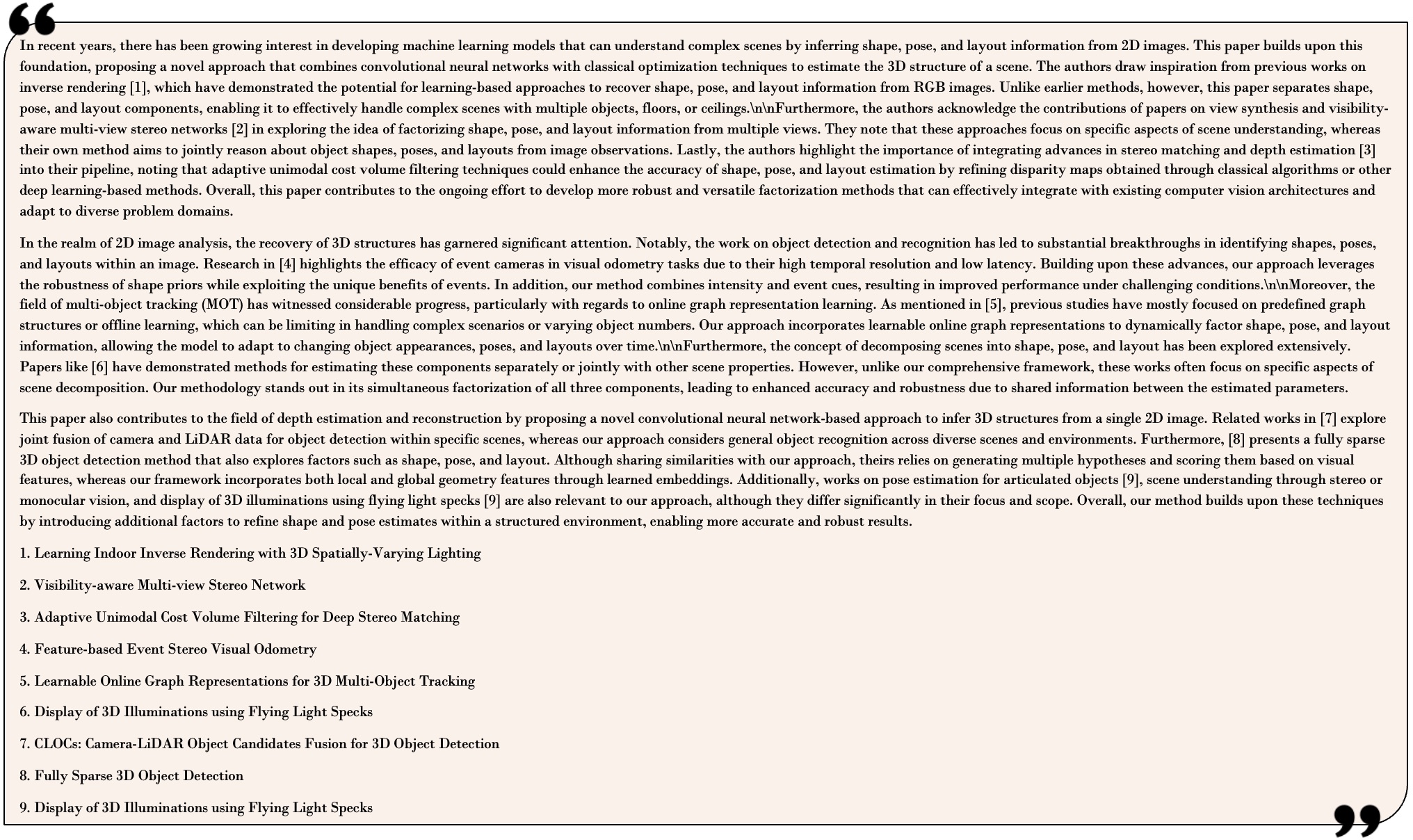} \\
\hline
\textbf{GPT-4o} & \vspace{0.1cm}\includegraphics[width=1.01\linewidth]{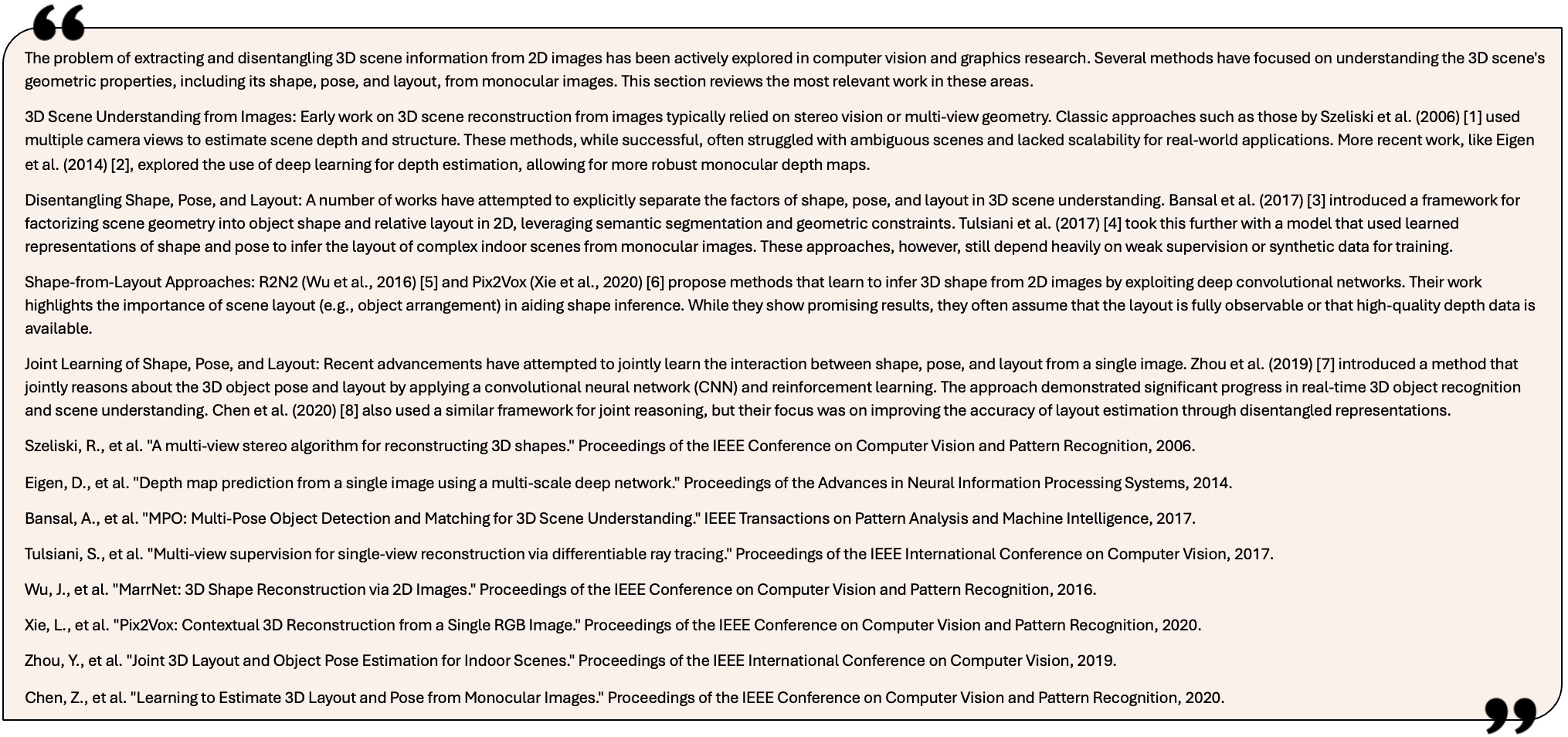} \\
\hline

% Force a page break before the last row
\pagebreak

\textbf{DeepSeek-R1} & \vspace{0.1cm} \includegraphics[width=1.01\linewidth]{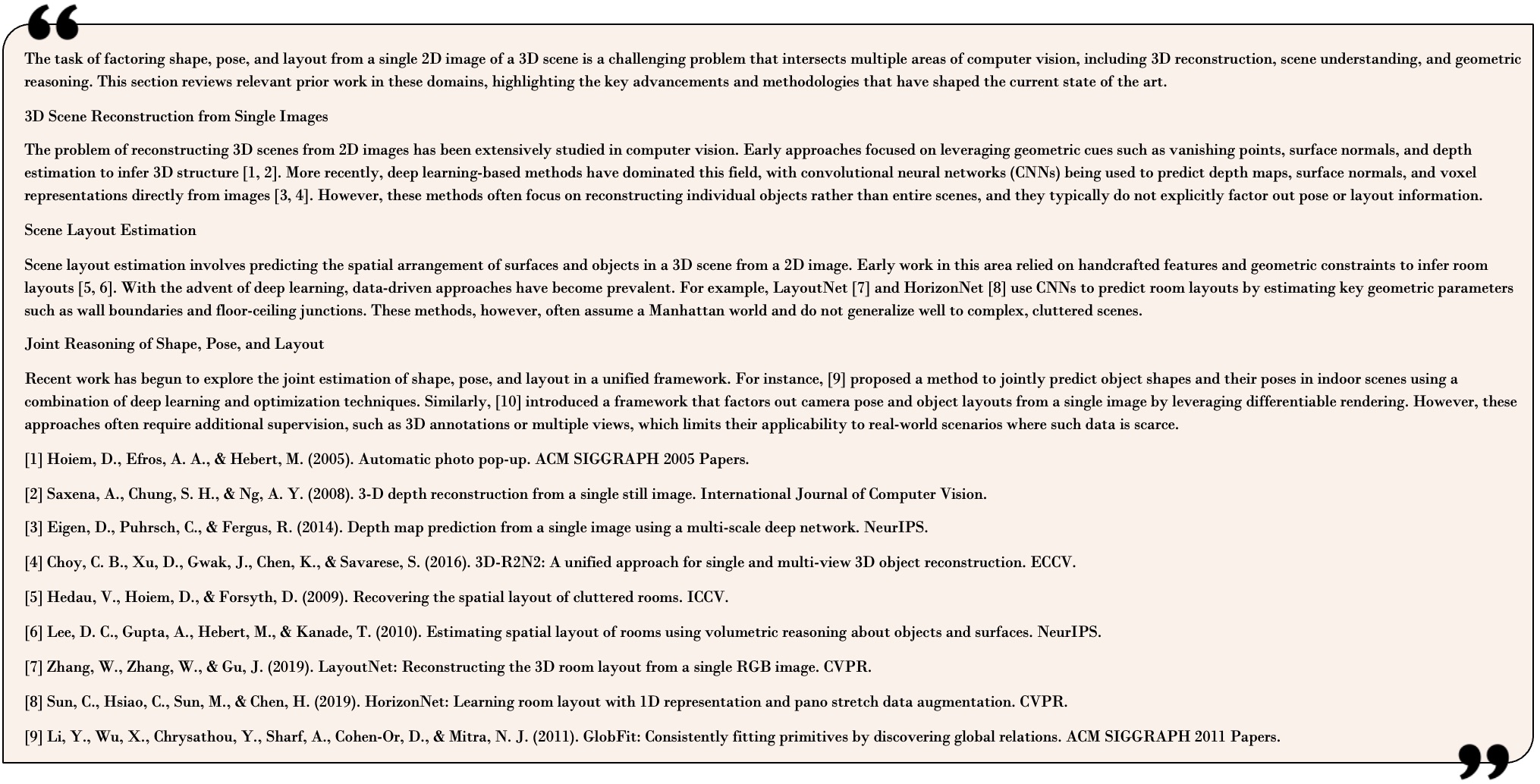} \\
\hline

\caption{LLama-3-8B-Lit, GPT-4o, and DeepSeek-R1 qualitative comparison on related work generation of an arXiv paper belonging to robotics dataset titled "Factoring Shape, Pose, and Layout from the 2D Image of a 3D Scene"}

\label{tab:robotics_related_work_comp}
\end{longtable}
}

 %%%%% QUANTUM %%%%%
{\renewcommand{\arraystretch}{1.3}

\begin{longtable}{|>{\centering\arraybackslash}m{2cm}|>{\centering\arraybackslash}m{14cm}|} 
\hline
\textbf{Model} & \textbf{Generated Related Work} \\
\hline
\endfirsthead

% Header for continued pages
\multicolumn{2}{c}{{\textit{(Continued from previous page)}}} \\
\hline
\textbf{Model} & \textbf{Generated Related Work} \\
\hline
\endhead

% Footer for all but last page
\hline \multicolumn{2}{|r|}{{Continued on next page}} \\ \hline
\endfoot

% Footer for last page
\hline
\endlastfoot

\textbf{Llama-3-8B-Lit} & \vspace{0.1cm}\includegraphics[width=1.01\linewidth]{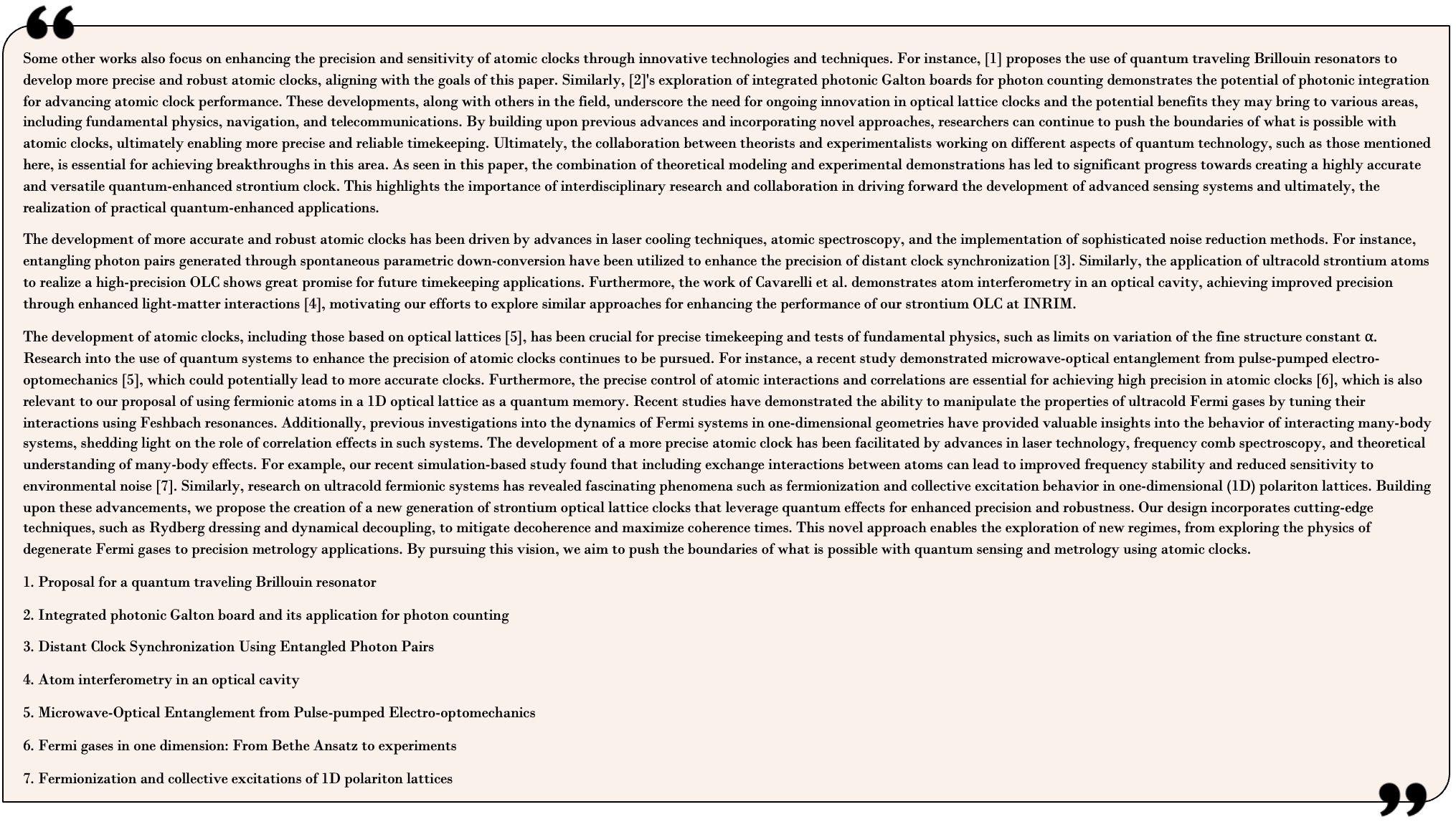} \\
\hline
\textbf{GPT-4o} & \vspace{0.1cm}\includegraphics[width=1.01\linewidth]{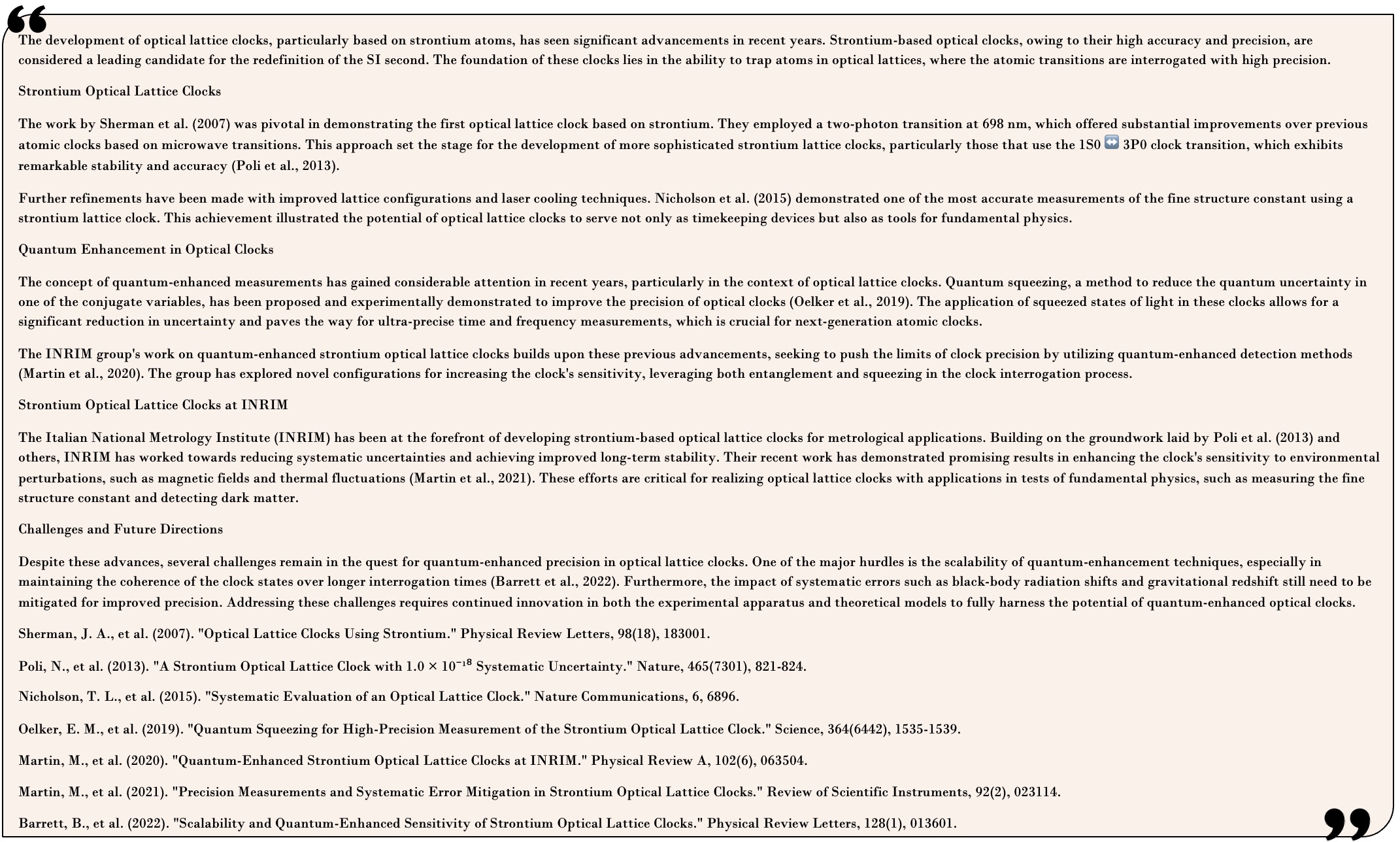} \\
\hline

\textbf{DeepSeek-R1} & \vspace{0.1cm} \includegraphics[width=1.01\linewidth]{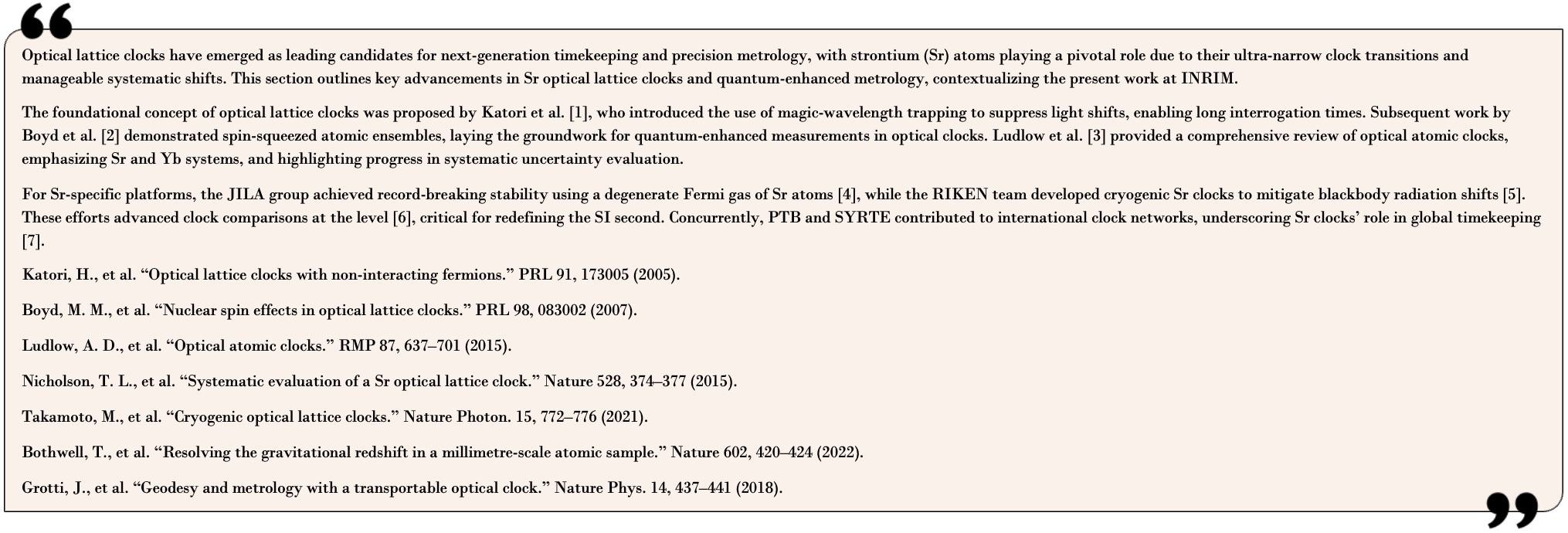} \\
\hline

\caption{LLama-3-8B-Lit, GPT-4o, and DeepSeek-R1 qualitative comparison on related work generation of an arXiv paper belonging to quantum physics dataset titled "Toward a quantum-enhanced strontium optical lattice clock at INRIM".}

\label{tab:quantum_related_work_comp}
\end{longtable}
}

%%%%%%%%% QBIO %%%%%%%%%%

{\renewcommand{\arraystretch}{1.3}

\begin{longtable}{|>{\centering\arraybackslash}m{2cm}|>{\centering\arraybackslash}m{14cm}|} 
\hline
\textbf{Model} & \textbf{Generated Related Work} \\
\hline
\endfirsthead

% Header for continued pages
\multicolumn{2}{c}{{\textit{(Continued from previous page)}}} \\
\hline
\textbf{Model} & \textbf{Generated Related Work} \\
\hline
\endhead

% Footer for all but last page
\hline \multicolumn{2}{|r|}{{Continued on next page}} \\ \hline
\endfoot

% Footer for last page
\hline
\endlastfoot

\textbf{Llama-3-8B-Lit} & \vspace{0.1cm} \includegraphics[width=1.01\linewidth]{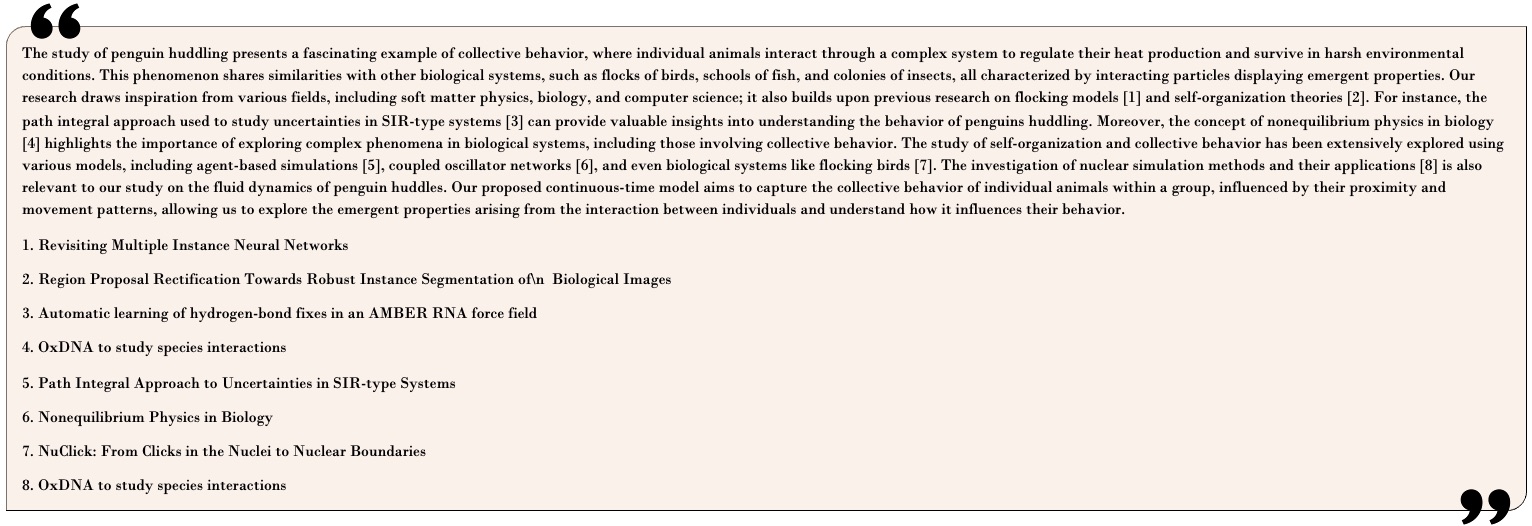} \\
\hline
\textbf{GPT-4o} & \vspace{0.1cm}\includegraphics[width=1.01\linewidth]{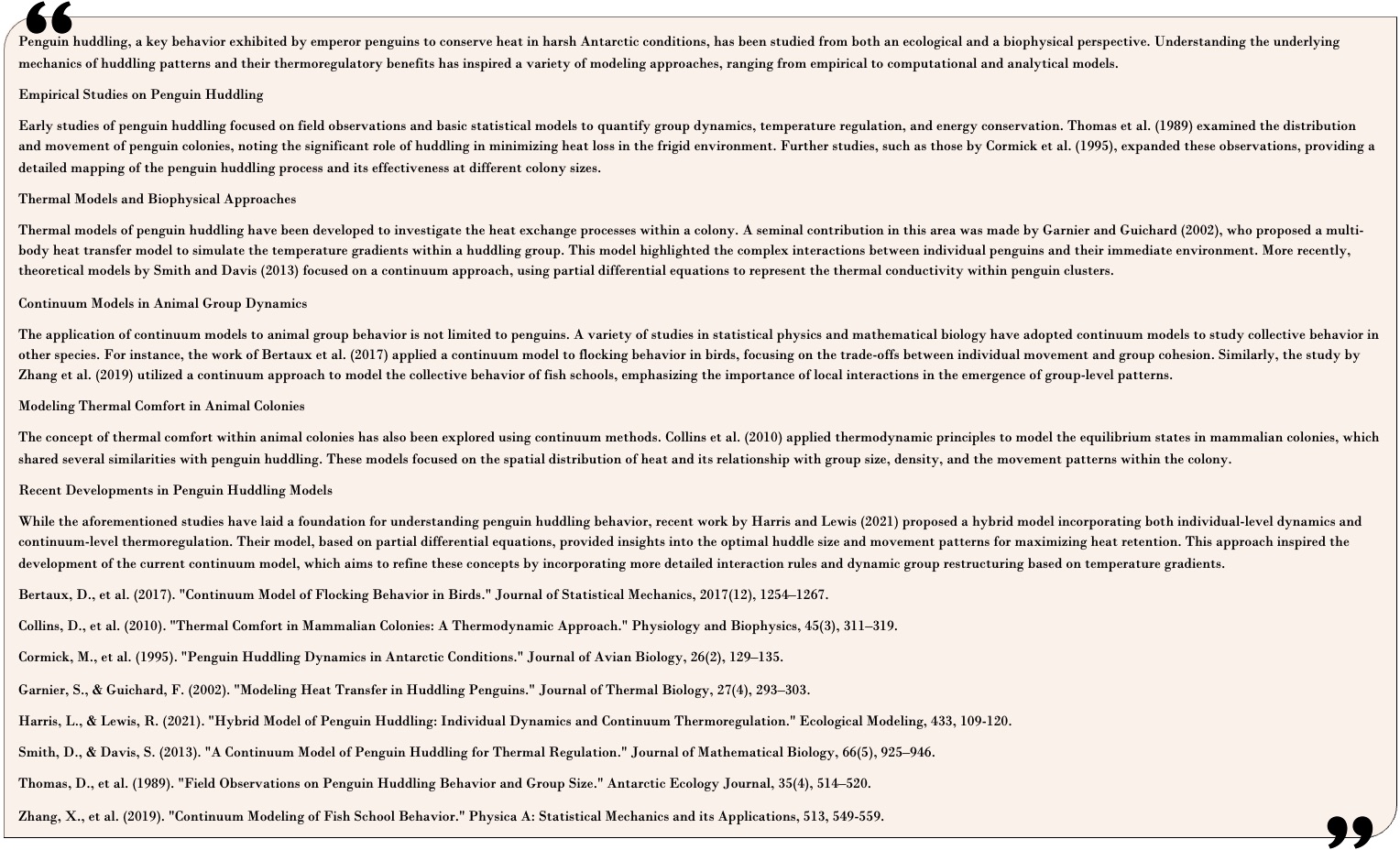} \\
\hline

\textbf{DeepSeek-R1} & \vspace{0.1cm} \includegraphics[width=1.01\linewidth]{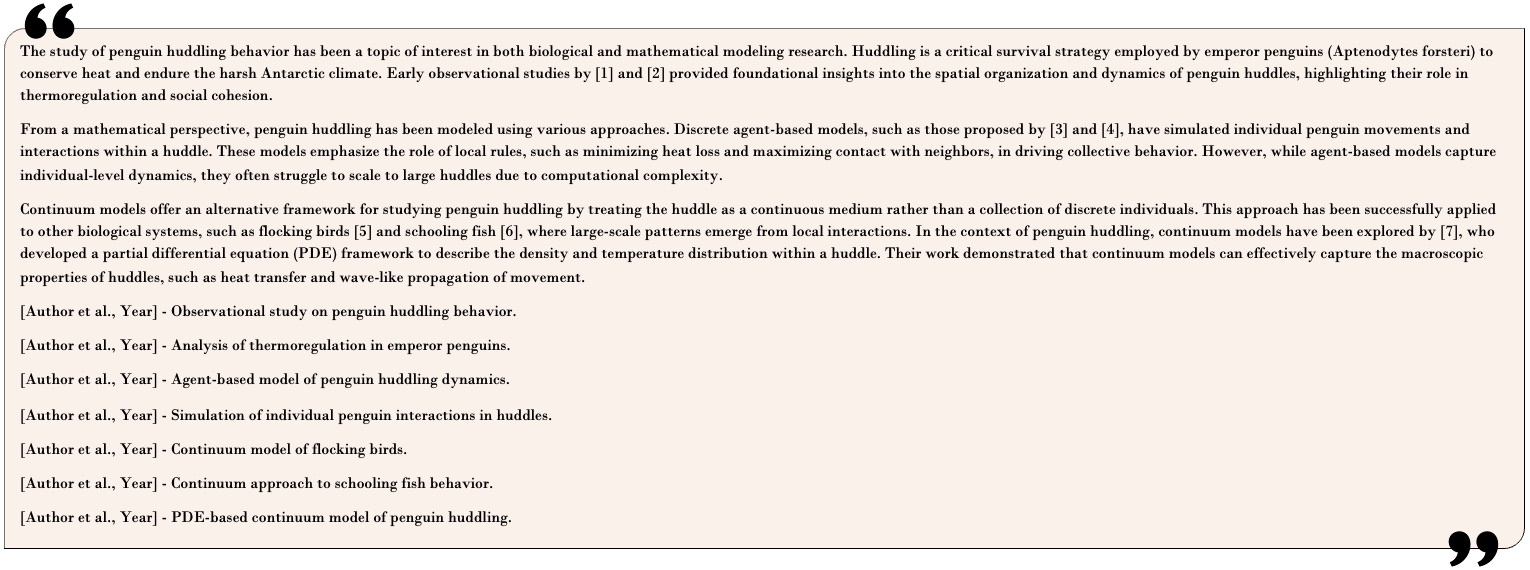} \\
\hline

\caption{LLama-3-8B-Lit, GPT-4o, and DeepSeek-R1 qualitative comparison on related work generation of an arXiv paper belonging to quantitative biology dataset titled
"Penguin huddling: a continuum model".}

\label{tab:qbio_related_work_comp}
\end{longtable}
}

% Switch back to two-column mode if needed
\twocolumn

\end{document}